\documentclass[twocolumn, aps, prd, superscriptaddress, preprintnumbers, eqsecnum, floatfix, nofootinbib]{revtex4}
\usepackage[colorlinks=true, pdfstartview=FitV, linkcolor=red, citecolor=blue, urlcolor=blue]{hyperref}
\usepackage{color}

\usepackage{multirow}
\usepackage{epsfig}
\usepackage{amsmath}

\usepackage{latexsym}
\usepackage{feynmp}

\usepackage[normalem]{ulem}  

\renewcommand\sout{\bgroup \color{red} \ULdepth=-.5ex \ULset}

\newcommand{\vect}[1]{\mathbf{#1}}

\newcommand{\sgn}{\mathrm{sgn}}
\newcommand{\Slash}[1]{\ooalign{\hfil/\hfil\crcr$#1$}}

\newcommand{\vp}{\vect{p}}
\newcommand{\vk}{\vect{k}}

\newcommand{\vl}{\vect{l}}

\newcommand{\vzero}{\vect{0}}

\newcommand{\vA}{\vect{A}}
\newcommand{\vE}{\vect{E}}
\newcommand{\vB}{\vect{B}}

\newcommand{\Tr}{\mathrm{Tr}}

\newcommand{\comment}[1]{}

\newcommand{\Nc}{N_c}
\newcommand{\Nf}{N_f}
\newcommand{\Cf}{C_F}
\newcommand{\nf}{n_F}
\newcommand{\nb}{n_B}
\newcommand{\qf}{q_f}
\newcommand{\Bf}{B_f}
\newcommand{\rf}{r^f}
\newcommand{\eL}{\epsilon^L}

\newcommand{\Tc}{{\mathrm T}_C}

\newcommand{\mf}{m_f}

\begin{document}

\preprint{RBRC-1206}

\title{Electrical Conductivity of Quark-Gluon Plasma in Strong Magnetic Fields}

\author{Koichi Hattori}
\email[]{koichi.hattori@riken.jp}
\affiliation{Physics Department and Center for Particle Physics and Field Theory, 
Fudan University, Shanghai 200433, China
}
\affiliation{RIKEN-BNL Research Center, Brookhaven National Laboratory, Upton,
 New York 11973-5000, U.S.A}

\author{Daisuke Satow}
\email[]{dsato@th.physik.uni-frankfurt.de}
\affiliation{Institut f\"ur Theoretische Physik, Johann Wolfgang Goethe-Universit\"at,
Max-von-Laue-Str. 1, D-60438 Frankfurt am Main, Germany
}

\date{\today}

\begin{abstract}
We compute the electrical conductivity of quark-gluon plasma in a strong magnetic field $B$ with quantum field theory at finite temperature using the lowest Landau level approximation. 
We provide the one-loop result arising from 
1-to-2 scattering processes whose kinematics are satisfied 
by the (1+1) dimensional fermion dispersion relation. 
Due to the chirality conservation, the conductivity diverges in the massless limit, and is sensitive to the value of the current quark mass. 
As a result, we find that the conductivity along the direction of the magnetic field 
is quite large compared with the value at $ B=0$, 
mainly because of the small value of the current quark mass. 
We show that the resummation of the ladder diagrams 
for the current-current correlator gives rise to only sub-leading contributions beyond the leading-log order, and thus verify our one-loop result at the leading-log accuracy. 
We also discuss possible implications for the relativistic heavy-ion collisions.
\end{abstract}

\maketitle

\section{Introduction}

The relativistic heavy-ion collision programs at RHIC and LHC have been 
providing successful results for the creation of quark-gluon plasma (QGP) at high temperature, 
and various properties of quantum chromodynamics (QCD) at the extreme condition have been investigated. 
It was also suggested that an extremely strong magnetic field ($B$) is induced in the noncentral collisions 
by the Ampere's circuital law~\cite{KMW,Bestimates} (see Ref.~\cite{Hattori:2016emy} for recent reviews). 
The magnitude of the magnetic field is estimated to be of the order or larger than the QCD scale 
 $\varLambda_{\text{QCD}}$ ($\varLambda_{\text{QCD}}^2 \lesssim eB$ with $e$ being 
the coupling constant in quantum electrodynamics). 
Since such a strong magnetic field has not been realized in other terrestrial experiments, 
the heavy-ion collisions provides us with a unique opportunity to investigate the properties of QCD matter 
at the high temperature and in the strong magnetic field.
Understanding the properties of QCD matter in strong magnetic fields 
can be also useful for neutron star and magnetar physics, 
where a strong magnetic field and a high density state are expected to be realized. 

In the recent years, the strong magnetic field induced by the heavy ion collisions 
have attracted a number of interests. The anomaly-induced transport, 
the so-called chiral magnetic effect \cite{KMW}, triggered not only intensive theoretical studies 
but also experimental efforts (see Refs.~\cite{Kharzeev:2013ffa, Skokov:2016yrj, Hattori:2016emy} for reviews). 
While significant progresses have been made, the interpretation of the experimental 
results appear to be still controversial due to the uncertainties 
such as the lifetime of the magnetic field, the distribution of the axial charges, 
the background of the experimental signal, etc. 
Therefore, deeper understanding of various aspects of the QGP in the strong magnetic field 
has been becoming important to achieve the consistent dynamical modelling.  

In this perspective, the transport coefficients are important quantities. 
Preceding studies addressed effects of the magnetic field on, e.g., 
the electrical conductivity \cite{Alford:2014doa,Buividovich:2010tn, Pu:2014fva, 
Satow:2014lia, Gorbar:2016qfh, Harutyunyan:2016rxm, Kerbikov:2014ofa, Nam:2012sg}, 
the shear viscosity \cite{Huang:2011dc,Tuchin:2013ie, Finazzo:2016mhm,Alford:2014doa}, 
the heavy-quark diffusion constant \cite{Fukushima:2015wck, Finazzo:2016mhm}, and 
the jet quenching parameter \cite{Li:2016bbh} by various methods and 
with different assumptions for the hierarchy of scales. 
Since there were several progresses also in the dynamical modelling of 
the anomalous charge separation \cite{Hongo:2013cqa}, 
the magnetohydrodynamics \cite{Roy:2015kma, Inghirami:2016iru, Pu:2016bxy}, 
and the Langevin dynamics for open heavy flavors \cite{Das:2016cwd}, 
it is an urgent task to compute the transport coefficients from the microscopic theories. 

In this paper, we focus on one of the most important transport coefficients 
in the magnetohydrodynamics, that is, the electrical conductivity of QGP. 
This quantity is interesting from the phenomenological point of view: 
If the conductivity is large enough, it is expected that the magnetic field 
induced by the heavy ion collision persists longer in time~\cite{
McLerran:2013hla, Tuchin:2013apa, Tuchin:2015oka, Hattori:2016emy}. 

Also, from the theoretical point of view, we will find 
a drastic change of the relaxation dynamics in the strong magnetic field limit. 
This change is originated from the Landau level quantization for the periodic cyclotron motion: 
The quark spectrum in the magnetic field (applied in the $ z$-direction) is discretized as 
\begin{eqnarray}
\epsilon_n = \sqrt{p_z^2 + \mf^2 + 2n \vert eB \vert }
\, ,
\end{eqnarray}
where $\mf$ is the current quark mass and $p_z$ is the $z$ component of the quark momentum. 
Therefore, the low-energy fermion dynamics is dominated by the (1+1)-dimensional ground state ($ n=0$), 
i.e., the lowest Landau level (LLL).  
This is the quantum system realized in the strong magnetic field limit 
where the magnitude of the magnetic field is 
much greater than the other energy scales of the system such as temperature. 
On the other hand, the electrically neutral gluons are 
not coupled to the magnetic field at the leading order in weak-coupling theories, 
so that they move in the three dimensions. 
Thus, we need to consider the transport process with 
the fermions moving in one dimension and the bosons moving in three dimensions. 
This is an intriguing system which is quite different from 
both the usual (3+1)-dimensional theory, and the (1+1)-dimensional theory 
where there is no dynamical gluonic degrees of freedom and the fermions suffer the confinement. 
In fact, the effect of the strong magnetic field opens novel 1-to-2 scattering processes 
(see Refs.~\cite{Tuchin:2010gx, Hattori:2012je} for the study at $ T=0$), 
which were forbidden by the kinematic reason when $B=0$. 
In addition, in the massless limit $( \mf = 0)$, the chirality conservation forbids the scattering process~\cite{Smilga:1991xa}, 
so that the conductivity, which diverges without scatterings, is expected to be very sensitive 
to $\mf$ even when the mass is quite small. 
It makes a striking contrast to the computations of the transport coefficients 
without a magnetic field~\cite{Jeon:1994if, Gagnon:2006hi,  Hidaka:2010gh, Wang:2002nba, Arnold:2000dr}, 
where we could safely neglect the current quark mass at the high temperature $T \gg \mf$. 
While the conductivity in weak magnetic fields 
has been evaluated by lattice QCD~\cite{Buividovich:2010tn}, AdS/CFT correspondence~\cite{Pu:2014fva}, 
and the Boltzmann equation~\cite{Satow:2014lia, Gorbar:2016qfh, Harutyunyan:2016rxm, Kerbikov:2014ofa}, 
this strong-field regime has not been explored.

We will evaluate the electrical conductivity in strong magnetic field 
at finite temperature and vanishing chemical potential. 
As discussed shortly, we use the lowest Landau level (LLL) approximation, 
and our calculation is performed at the leading-log accuracy. 
It is known that the transport coefficients, including the electrical conductivity, 
can be consistently obtained from the kinetic equation~\cite{Arnold:2000dr} 
and the diagrammatic method~\cite{Jeon:1994if, Gagnon:2006hi, Hidaka:2010gh, Wang:2002nba}. 
However, in the presence of the magnetic field, 
the ordinary kinetic equation will not be directly applicable due to 
the quantum nature of the Landau levels, 
and one needs to elaborate the construction of kinetic equation. 
In an accompanying paper~\cite{another-paper}, 
one can find the formulation of an effective kinetic equation 
and the evaluation of the conductivity beyond leading-log accuracy. 
In this paper, starting out from quantum field theory, 
we show that a consistent conclusion is drawn by using the diagrammatic method, 
and briefly discuss an equivalence to the kinetic equation in Appendix~\ref{app:linear-Boltzmann}.

This paper is organized as follows:
In the next section, we introduce how to evaluate the electrical conductivity in the real time formalism.
By performing one-loop order analysis, we obtain the expression of the conductivity written in terms of the quark damping rate, and explicitly evaluate the damping rate generated by the 1-to-2 scatterings in Sec.~\ref{sec:one-loop}.
Section~\ref{sec:results} is devoted to explaining the features of the result for the conductivity. 
In Sec.~\ref{sec:ladder}, we discuss the resummation of the ladder diagrams. 
We briefly discuss possible implications of our results for the heavy ion collision experiments in Sec.~\ref{sec:heavyion}.
In Sec.~\ref{sec:summary}, we summarize this paper and give a few concluding remarks.
In the four Appendices, we discuss the gauge-fixing independence of our result, 
the integration range with respect to the energy of the scattering particle in the 1-to-2 scattering process, 
consistency of our diagrammatic scheme to the Ward-Takahashi identity, 
and the equivalence of our scheme to the approach with kinetic equation, respectively.

Prior to going into explicit computations, 
we would like to discuss the characteristic energy scales involved in the problem, 
and specify our hierarchy assumed throughout this paper. 
In the analysis below, a few characteristic energy scales appear:
The largest energy scale, $\sqrt{eB}$, is due to the magnetic field.
Because we work in the strong magnetic field regime, we assume that it is much larger than the temperature, $\sqrt{eB}\gg T$. 
This condition justifies the usage of the LLL approximation, i.e., neglect of the higher Landau levels.
We also have the current quark mass ($\mf$).
In most calculation at QGP phase, this quantity has been neglected because it is of order $\sim 1$MeV while $T\sim 100$MeV.
When infrared divergence appears, it was regulated by thermal masses of quarks and gluons.
However, in the LLL approximation, we cannot neglect $\mf$ because the scattering processes are forbidden if $\mf=0$ due to the chirality conservation~\cite{Smilga:1991xa}, as we will discuss later.
On the other hand, the gluon also dynamically gets a screening mass ($M$), which is of order $g\sqrt{eB}$ ($g$: QCD coupling constant)~\cite{Fukushima:2011nu, Fukushima:2015wck}.
Because we are interested in the case that finite-$T$ effect is significant, 
we consider the case of $\mf, M\ll T$, where the quarks and gluons are thermally well excited.
Summarizing, we work in the regime $\mf, M\ll T\ll \sqrt{eB}$.
As for the ordering of $\mf$ and $M$, we consider both of the cases:
$\mf\gg M$ and $\mf\ll M$.

\section{Electrical Conductivity in real time formalism}

In this section, we introduce how to evaluate the electrical conductivity in the real time formalism.
We begin with formal introduction of the electrical conductivity.
Consider the situation that, the system is initially at equilibrium whose temperature is $T$ in magnetic field $\vB$, and then external electric field $\vE$ disturbs the system and induces electromagnetic current $j^\mu$.
Due to the linear response theory, the retarded current correlator 
\begin{align}
\varPi^{R \mu\nu}(x)\equiv i\theta(x^0)
\langle [j^\mu(x), j^\nu(0)]
\rangle,
\end{align}
 determines the induced current in the momentum space: 
\begin{align}
j^\mu(p)&= -\varPi^{R \mu\nu }(p)A_\nu(p),
\end{align}
where $A_\nu$ is a vector potential that creates $\vE$.
When $\vE$ is homogenous in space, we have $\vp=0$ and $\vE=i\omega\vA$, and thus 
$j^i(\omega)= \varPi^{R ij }(\omega) E^j(\omega) /(i\omega)$.
By taking $\omega\rightarrow 0$ limit, we have $j^i=\sigma^{ij}E^j$, where we have introduced the DC conductivity tensor, 
\begin{align}
\sigma^{ij}
\equiv \lim_{\omega\rightarrow 0}\frac{\varPi^{R ij }(\omega)}{i\omega} .
\end{align}
Thus, the DC conductivity can be evaluated by calculating the current correlator in low energy limit at zero momentum, whose expression is called Kubo formula.

\begin{figure}[t!] 
\begin{center}
\includegraphics[width=0.3\textwidth]{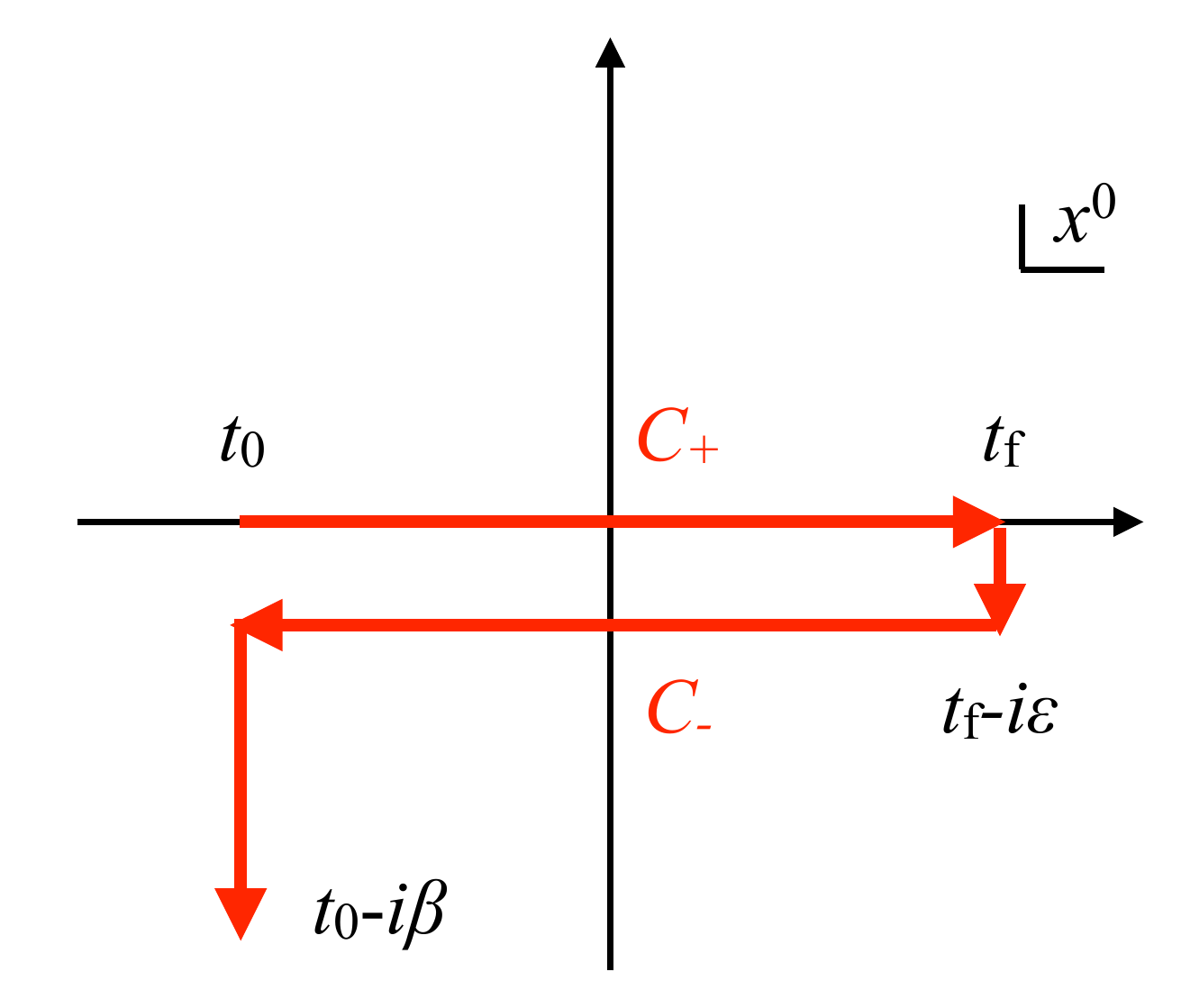} 
\caption{The contour in complex time plane.
The part $C_+$ is on the real axis and $C_-$ is below that axis by $\epsilon$.
} 
\label{fig:contour} 
\end{center} 
\end{figure}

This expression can be rewritten as follows in the real time formalism:
By using $\nb(\omega)\simeq T/\omega$ for $\omega\ll T$, where $\nb(\omega)\equiv [e^{\beta \omega}-1]^{-1}$ is the Bose distribution function with $\beta\equiv 1/T$, we have 
\begin{align}
\label{eq:sigma-Pi12}
\sigma^{ij}
= \frac{\beta}{2}
\varPi^{ij}_{12}(\omega\rightarrow 0) ,
\end{align}
where $\varPi^{\mu\nu}_{12}(x)\equiv \langle \Tc j^\mu_1(x) j^\nu_2(0) \rangle=  \langle j^\nu(0) j^\mu(x)  \rangle$ and $\rho^{\mu\nu}(p)\equiv 2{\text {Im}}\varPi^{R \mu\nu }(p)$ is the spectral function of the current, that satisfies $\varPi^{\mu\nu}_{12}(p)=\nb(\omega)\rho^{\mu\nu}(p)$.
Here we have introduced the contour in complex time drawn in Fig.~\ref{fig:contour}, where the limits $t_0\rightarrow -\infty$ and $t_f\rightarrow \infty$ are taken.
$\Tc$ is an ordering operator on this contour and $j^\mu_{1/2}$ is a current operator whose time belongs to $C_{1/2}$.
For more detail of the real time formalism, see~\cite{lebellac, Blaizot:2001nr}.

Here we write the current correlator in terms of quark field for evaluation.
The current operator is defined as 
\begin{align}
j^\mu(x)&\equiv e\sum_f {\qf} \overline{\psi}_f(x) \gamma^\mu \psi_f(x),
\end{align}
where $f$ is an index for flavor, $\qf$ is a EM charge for the quark, and $\psi_f$ is a quark operator.
In the LLL approximation, the quark field reads~\cite{Hattori:2015aki}
\begin{align}
\psi_f(x)
&= \int_{p_L, p^2}
e^{-i (p_L\cdot x_L - p^2 x^2)}
{\cal P}^f_+
\chi^f_{p^2}(p_L) {\cal H}(x^1-r^f_{p^2}),
\end{align}
where we have adopted the Landau gauge considered the case that $\vB$ is along $z$-axis, in which $A^2_{\text{ext}}=Bx^1$ with $A^\mu_{\text{ext}}$ the vector potential that yields the magnetic field.
We have also introduced $d^2 p_L\equiv d p^0 dp^3$, $p_L\equiv (p^0,0,0, p^3)$, $\int_p\equiv \int dp/(2\pi)$, $\rf_{p^2}\equiv -p^2/\Bf$, $\Bf\equiv e\qf B$, and $\chi^f_{p^2}(p_L)$ is the quark operator at the LLL.
We note that $p^2$ here does not mean the square of four-vector $p^\mu$, $(p^0)^2-\vp^2$, but the $y$-component of $p^\mu$.
${\cal P}^f_\pm\equiv (1\pm \sgn(\Bf)i\gamma^1\gamma^2)/2$ is a projection operator into a state with spin aligning with $\vB$.
${\cal H}(x-r^f)\equiv [|\Bf|/\pi]^{1/4}\exp[-|\Bf|(x-r^f)^2/2]$ is the normalized harmonic oscillator function coming from quark wave function in transverse plane at the LLL.
In this approximation, the current operator becomes
\begin{align}
\begin{split}
j^\mu(x)
&= e\sum_f {\qf} 
\int_{p^2, p_L, k^2, k_L} e^{-ix_L\cdot(k_L-p_L)}e^{ix^2(k^2-p^2)} \\
&~~~\times{\cal H}(x^1-r^f_{p^2}){\cal H}(x^1-r^f_{k^2})
\overline{\chi}^f_{p^2}(p_L) \gamma^\mu \chi^f_{k^2}(k_L).
\end{split}
\end{align}
The current correlator can be written in terms of four-point function of quark:
\begin{widetext}
\begin{align}
\label{eq:Pimunu-A}
\begin{split} 
\varPi^{\mu\nu}_{12}(p=0)
&= e^2\sum_{f,f'} q_f q_{f'} 
\int_{k^2,k_L, q^2,q_L, l^2,l_L,r^2,r_L}
(2\pi)\delta(k^2-q^2) (2\pi)^2 \delta^{(2)}(k^L-q^L) 
(2\pi)\delta(l^2-r^2) (2\pi)^2 \delta^{(2)}(l^L-r^L) \\
&~~~\times \left[\int dx^1 {\cal H}(-x^1-r^{f'}_{l^2})  {\cal H}(-x^1-r^{f'}_{r^2}) \right]
{\cal H}(-r^{f}_{k^2}){\cal H}(-r^{f}_{q^2}) 
 \left\langle \Tc \overline{\chi}^f_{1 k^2}(k_L) \gamma^\mu \chi^f_{1 q^2}(q_L)
\overline{\chi}^{f'}_{2 l^2}(l_L) \gamma^\nu \chi^{f'}_{2 r^2}(r_L)
\right\rangle \\
&= e^2\sum_{f,f'} q_f q_{f'} 
\int_{k^2,k_L,  l^2,l_L} 
\left[ {\cal H}(-r^{f}_{k^2}) \right]^2
 \left\langle \Tc \overline{\chi}^f_{1 k^2}(k_L) \gamma^\mu \chi^f_{1 k^2}(k_L)
\overline{\chi}^{f'}_{2 l^2}(l_L) \gamma^\nu \chi^{f'}_{2 l^2}(l_L)
\right\rangle,
\end{split}
\end{align}
\end{widetext}
where we have used $\int dx [{\cal H}(x)]^2=1$.
When we use the symmetry in the color space and neglect the flavor changing process, which will be justified in the analysis later since we will consider only the ladder diagrams, the four-point function above has the structure $ \left\langle \Tc \overline{\chi}^f_{1 k^2}(k_L) \gamma^\mu \chi^f_{1 k^2}(k_L)\overline{\chi}^{f'}_{2 l^2}(l_L) \gamma^\nu \chi^{f'}_{2 l^2}(l_L)\right\rangle=\Nc \delta_{ff'}G^{\mu\nu f}_{1122}(k,k,l,l) $, with the four-point function $G^{\mu\nu f}_{1122}(k,k,l,l)\equiv \left\langle \Tc \overline{\chi}^f_{1 k^2}(k_L) \gamma^\mu \chi^f_{1 k^2}(k_L)\overline{\chi}^{f}_{2 l^2}(l_L) \gamma^\nu \chi^{f}_{2 l^2}(l_L) \right\rangle$.
Here the color summation in $G^{\mu\nu f}_{1122}$ has been done.
We further assume that $\int_{l^2}G^{\mu\nu f}_{1122}(k,k,l,l) $ does not depend on $k^2$, which also will be justified later.
Then, $\varPi^{\mu\nu}_{12}$ can be written as 
\begin{align}
\label{eq:Pi-12-G-1122}
\begin{split} 
\varPi^{\mu\nu}_{12}(p=0)
&= e^2\sum_{f} (q_f)^2 \Nc \frac{|B_f|}{2\pi}
\int_{k^2,k_L, l_L} 
 G^{\mu\nu f}_{1122}(k,k,l,l),
\end{split}
\end{align}
where we have used $\int_p [{\cal H}(-r^f_p)]^2=|B_f|/2\pi$.

For later convenience, we move to r/a basis.
By introducing $\chi^r\equiv (\chi^1+\chi^2)/2$ and $\chi^a\equiv \chi^1-\chi^2$, the four-point function can be written as 
\begin{align}
\begin{split}
G_{1122}
&= G_{rrrr}
+\frac{1}{2}\left(G_{arrr}+G_{rarr}-G_{rrar}-G_{rrra}\right) \\
&~~~+\frac{1}{4}(G_{aarr}-G_{arar}-G_{arra}\\
&~~~-G_{raar}-G_{rara}+G_{rraa}) \\
&~~~+\frac{1}{8}\left(-G_{aaar}-G_{aara}+G_{araa}+G_{raaa}\right),
\end{split}
\end{align}
where we have omitted the Lorentz/flavor indices for simplicity, introduced $G_{ijkl}\equiv\langle\Tc \overline{\chi}_{i} \gamma^\mu \chi_{j}\overline{\chi}_{k} \gamma^\nu \chi_{l}\rangle$ with $i, j, k, l=r$ or $a$, and used $G_{aaaa}=0$.
By using the generalized fluctuation-dissipation theorem~\cite{Wang:2002nba, Wang:1998wg, Gervais:2012wd}, this expression can be rewritten as 
\begin{align}
\label{eq:G1122-ra}
\begin{split}
G_{1122}
&= \alpha_1 G_{aarr}
+ \alpha_2 G_{aaar}+ \alpha_3 G_{aara}+ \alpha_4 G_{araa}\\
&~~~+ \alpha_5 G_{raaa}
+ \alpha_6 G_{arra}+ \alpha_7 G_{arar}\\ 
&~~~+\beta_1 \overline{G}^*_{aarr}
+\beta_2 \overline{G}^*_{aaar}+\beta_3 \overline{G}^*_{aara}+\beta_4 \overline{G}^*_{araa} \\
&~~~+\beta_5 \overline{G}^*_{raaa}+\beta_6 \overline{G}^*_{arra}+\beta_7 \overline{G}^*_{arar},
\end{split}
\end{align}
where we have introduced the Fermi distribution function $\nf(E)\equiv[e^{E/T}+1]^{-1}$ and $\alpha_1=\beta_1\equiv \nf(k^0)[1-\nf(k^0)]$.
We do not write the other coefficients explicitly because they will be found to be irrelevant to the leading-order calculation.
The bar above $G$ means the interchange between the quark and the anti-quark:
$ \overline{G}^{\mu\nu}_{ijkl}\equiv \langle\Tc \overline{\chi}_{j} \gamma^\mu \chi_{i}\overline{\chi}_{l} \gamma^\nu \chi_{k}\rangle$.
Neglecting the irrelevant terms in Eq.~({\ref{eq:G1122-ra}}) 
and using $G_{aarr}(k,k,l,l)=\overline{G}_{aarr}(k,k,l,l)$, which can be shown by using the definition of $\overline{G}$, we get
\begin{align}
\begin{split}
G_{1122} 
&=2 \nf(k^0)[1-\nf(k^0)] {\text{Re}} G_{aarr}
+{\text{(other terms)}}.
\end{split}
\end{align}
It makes Eq.~(\ref{eq:sigma-Pi12}) as
\begin{align}
\label{eq:Pi-12-G-aarr}
\begin{split}
\sigma^{ij}
&= e^2\beta\sum_{f} (q_f)^2 \Nc \frac{|B_f|}{2\pi}
\int_{k^2,k_L, l_L} \\
&~~~\times \nf(k^0)[1-\nf(k^0)]
{\text{Re}} G^{ij f}_{aarr}(k,k,l,l),
\end{split}
\end{align}
by using Eq.~(\ref{eq:Pi-12-G-1122}).

\section{One-loop analysis}
\label{sec:one-loop}

In this section, we examine the four-point function in the LLL approximation, 
which is necessary for computing the electrical conductivity. 
While we work at the one-loop order, we use the dressed quark propagator 
in which the quark damping rate is resummed.
We also explicitly evaluate the quark damping rate 
taking into account the 1 to 2 scattering in both of the two cases, $\mf\gg M$ and $\mf\ll M$.

\subsection{Four-point function of quark}

\begin{figure}[t!] 
\begin{center}
\includegraphics[width=0.2\textwidth]{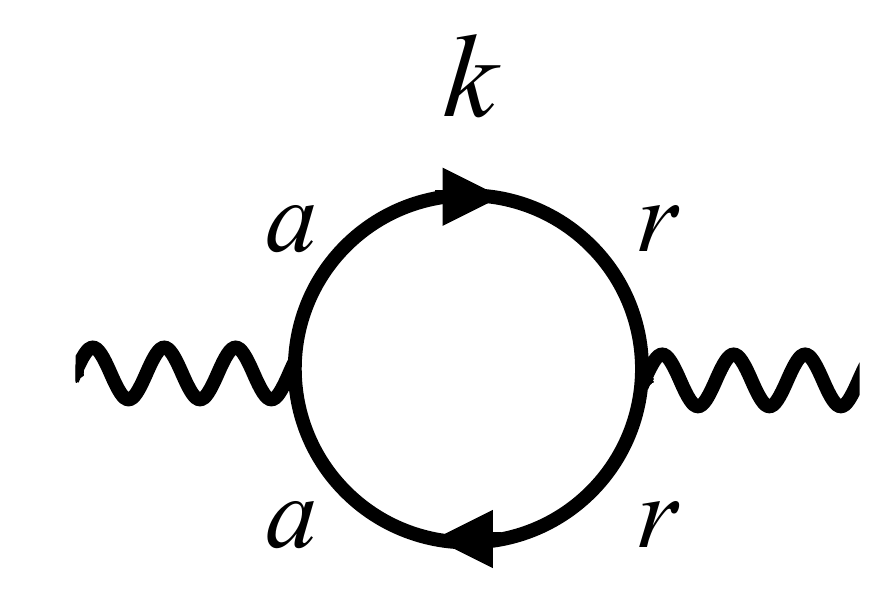} 
\caption{The current correlator $G^{\mu\nu f}_{aarr}(k,k,l,l)$ at one-loop level.
The solid line represents quark propagator.
} 
\label{fig:oneloop-Pi} 
\end{center} 
\end{figure}

We evaluate the four-point function at the one-loop approximation.
By using Wick's theorem, the four-point function becomes 
\begin{align} 
\label{eq:Wick}
\begin{split}
G^{\mu\nu f}_{aarr}(k,k,l,l)
&= -(2\pi)\delta(k^2-l^2) (2\pi)^2 \delta^{(2)}(k_L-l_L)\\
&~~~\times \Tr[\gamma^\mu S^{ar}(k_L)\gamma^\nu S^{ra}(k_L) ],
\end{split} 
\end{align}
where we have introduced the quark propagators in r/a basis, $S^{ij}(p_L)\equiv \langle \Tc \chi^i_{p^2}(p_L)\overline{\chi}^j_{p^2}(p_L) \rangle$, where $i,j=r,a$ and omitted flavor indices for simplicity.
We note that this quantity becomes independent of $k^2$ after integrating over $l^2$, 
so the assumption we made above Eq.~(\ref{eq:Pi-12-G-1122}) is justified.
It also implies that the $k^2$ integral in the right-hand side of Eq.~(\ref{eq:Pi-12-G-1122}) can be trivially performed thanks to $\delta(k^2-l^2)$, so that there remain only $k_L$ and $l_L$ integrals.
This property manifests the gauge invariance.
The other terms with different indices in Eq.~(\ref{eq:G1122-ra}) vanish because $S^{aa}=0$. 
This contribution corresponds to the one-loop diagram drawn in Fig.~\ref{fig:oneloop-Pi}. 
As we will see later, $S(k_L)$ is proportional to $(\Slash{k}_L+\mf){\cal P}_+$, so $G^{\mu\nu}$ vanishes when $\mu$, $\nu=\perp$ (1 or 2) because ${\cal P}_+\gamma^\perp {\cal P}_+=\gamma^\perp {\cal P}_-{\cal P}_+=0$. 
Therefore, we consider the case of $\mu$, $\nu=0$ or $3$ from now on.

If we naively use the quark propagator in the free limit in the expression above, we would have a divergence, which is called pinch singularity~\cite{Jeon:1994if, Gagnon:2006hi, Hidaka:2010gh, Wang:2002nba}.
To regulate this singularity, one needs to resum the damping rate of the quark ($\xi_p$). 
The resummed quark retarded/advanced propagator ($S^{R/A}=iS^{ra/ar}$) at the LLL approximation reads 
\begin{align} 
\label{eq:S-spinorstr}
S^{R/A}(p)
&= (\Slash{p}_L+\mf) {\cal P}_+ \varDelta^{R/A}_S(p),
\end{align}
where $\varDelta^{R/A}_S(p)\equiv -[p^2_L-\mf^2\pm 2i\xi_p p^0]^{-1}$, with $p^2_L\equiv (p^0)^2-(p^3)^2$ and $\mf$ is the current quark mass. 
The free part of the expression above is given in Ref.~\cite{Hattori:2015aki}, for example. 
We note that the modification of the quark mass $\mf$ due to the interaction effect is not necessary at the leading-order calculation, because the modification to $\mf$ is suppressed by the factor of $g^2\mf/T$, at most~\cite{Elmfors:1995gr}. 
This point is quite different from the $B=0$ case, where the quarks move in three-dimension so that the scattering process is finite even at $\mf=0$, and thus the thermal mass is independent from $\mf$ (it is of order $gT$~\cite{Klimov:1981ka}) when $T$ is large enough.
By using this expression, Eq.~(\ref{eq:Wick}) becomes
\begin{align} 
\begin{split}
G^{\mu\nu f}_{aarr}(k,k,l,l)
&= (2\pi)\delta(k^2-l^2) (2\pi)^2 \delta^{(2)}(k_L-l_L)\\
&~~~\times  \frac{k^\mu_L k^\nu_L}{\xi_k k^0} \rho_S(k_L),
\end{split} 
\end{align}
where we have evaluated the trace by using 
$\Tr[(\Slash{k}_L+\mf){\cal P}_+\gamma^\mu (\Slash{k}_L+\mf){\cal P}_+ \gamma^\nu]= 4k^\mu_L k^\nu_L$, which is obtained by using the on-shell condition $k^2_L=\mf^2$.
 We have also introduced the spectral function of the quark related to $\varDelta^{R/A}$, $\rho_S(p_L)\equiv -i[\varDelta^{R}_S(p_L)-\varDelta^{A}_S(p_L)]$, which satisfies $\varDelta^{R}_S(p_L)\varDelta^{A}_S(p_L)=\rho_S(p_L)/(4\xi_p p^0)$.

Using this resummed propagator, Eq.~(\ref{eq:Pi-12-G-aarr}) for $i=j=3$ 
becomes 
\begin{align}
\label{eq:Pi-damping}
\begin{split}
\sigma^{33}
&= \frac{\beta}{2}e^2\sum_{f} (q_f)^2 \Nc \frac{|B_f|}{2\pi}
\int\frac{dk^0 dk^3}{\pi} \\
&~~~\times \nf(k^0)[1-\nf(k^0)] 
\frac{(k^3)^2}
{\xi_k |k^0|} \delta(k^2_L-\mf^2),
\end{split}
\end{align}
where we have used the approximation 
\begin{align}
\label{eq:free-quark-spectrum}
\rho_S(k_L)\simeq (2\pi)\sgn(k^0)\delta(k^2_L-\mf^2).
\end{align}
We see that Eq.~(\ref{eq:Pi-damping}) is proportional to $\xi^{-1}_k$, so it diverges when $\xi_k\rightarrow 0$.
This is the pinch singularity, and physically it corresponds to the fact that the conductivity diverges when the quark do not scatter with other particles so that the mean free path becomes infinitely large.

\subsection{Quark damping rate}

\begin{figure}[t!] 
\begin{center}
\includegraphics[width=0.45\textwidth]{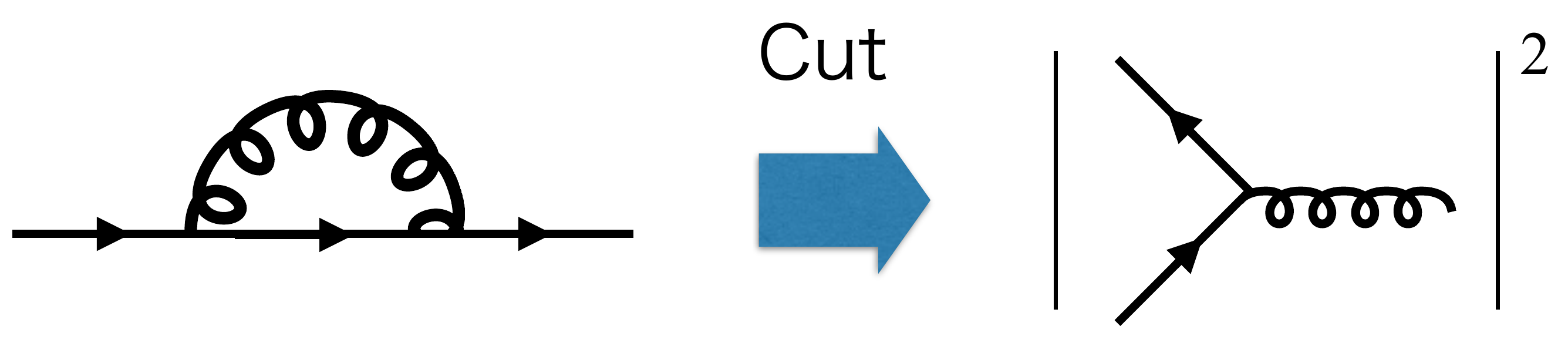} 
\caption{The quark self-energy at one-loop level (left panel) and the matrix element of the 1 to 2 scattering (right panel), which corresponds to the imaginary part of the self-energy.
The curly line represents gluon propagator.
} 
\label{fig:oneloop-Sigma} 
\end{center} 
\end{figure}

We need to evaluate the quark damping rate for proceeding the calculation.
The contribution from the one-loop diagram, which is drawn in the left panel of Fig.~\ref{fig:oneloop-Sigma}, is as follows: 
\begin{align}
\label{eq:ImSigma-oneloop}
\begin{split}
{\text {Im}}\varSigma^R(k_L)
&= \frac{g^2\Cf}{2} \int_{l}
\gamma^L_\mu(\Slash{l}_L+\mf){\cal P}_+ \gamma^L_\nu \rho^{\mu\nu}_{D}(k-l) \rho_S(l_L)\\
&~~~\times \left[R^f(\vk_\perp-\vl_\perp)\right]^2 [\nf(l^0)+\nb(l^0-k^0)],
\end{split} 
\end{align}
where we have introduced $k^\mu_\perp\equiv (0,k^1,k^2,0)$, $\Cf\equiv (\Nc^2-1)/(2\Nc)$, the form factor $R^f(\vp_\perp)\equiv \exp[-\vp^2_\perp/(4|\Bf|)]$, and the gluon spectral function $\rho^{\mu\nu}_{D}(k)\equiv 2{\text{Im}}D^{R \mu\nu}(k)$ with $D^{R \mu\nu}(k)$ the retarded gluon propagator.
We note that we have used the Ritus basis~\cite{Hattori:2015aki, Ritus:1972ky}, in which the momentum of the form factor is that of the gluon instead of the quark, and the Schwinger phases were canceled and thus did not appear in the expression above.
Again, this absence of the phases is a manifestation of the gauge invariance.
It yields the damping rate
\begin{align}
\label{eq:damping}
\begin{split}
\epsilon^L_k \xi_k
&= -\frac{1}{2}
\Tr \left[(\Slash{k}_L+\mf){\text {Im}}\varSigma^R(k_L)|_{k^0=\epsilon^L_k} \right]\\
&=- \frac{g^2\Cf}{4} \int_{l}
\Tr \left[(\Slash{k}_L+\mf)
\gamma^L_\mu(\Slash{l}_L+\mf){\cal P}_+ \gamma^L_\nu \right] \\
&~~~\times\rho^{\mu\nu}_{D}(k-l) \rho_S(l_L)\\
&~~~\times \left[R^f(\vk_\perp-\vl_\perp)\right]^2 [\nf(l^0)+\nb(l^0-k^0)]|_{k^0=\epsilon^L_k},
\end{split}
\end{align}
where we have introduced $\epsilon^L_k\equiv \sqrt{(k^3)^2+\mf^2}$, which is the on-shell energy of the quark with longitudinal momentum $k^3$.

We need to know the gluon propagator for evaluating the damping rate.
In the LLL approximation, the self-energy of the gluon coming from the quark loop 
has the tensor structure~\cite{Fukushima:2015wck} $\varOmega^{\mu\nu}(k)=\varOmega_{\parallel}(k)P^{\mu\nu}_\parallel(k_L)$, where $P^{\mu\nu}_\parallel(k_L)\equiv -[g^{\mu\nu}_L-k^\mu_L k^\nu_L/k^2_L]$ with $g^{\mu\nu}_L\equiv {\text {diag}}(1,0,0,-1)$. 
The self-energy coming from the gluon/ghost loop is much smaller, so we neglect it in this work.
The resultant gluon retarded propagator is, in the covariant gauge~\cite{Hattori:2012je, Fukushima:2015wck},
\begin{align}
\label{eq:gluon-propagator}
\begin{split}
D^{R \mu\nu}(k)
&=-\frac{P^{\mu\nu}_\parallel(k_L)}{k^2+i\epsilon k^0-\varOmega_{\parallel}(k)}
-\frac{P^{\mu\nu}_0(k)-P^{\mu\nu}_\parallel(k_L)}{k^2+i\epsilon k^0} \\
&~~~+\alpha\frac{k^\mu k^\nu}{(k^2+i\epsilon k^0)^2},
\end{split}
\end{align}
where $\alpha$ is a gauge-fixing parameter and $P^{\mu\nu}_0(k)\equiv -[g^{\mu\nu}-k^\mu k^\nu/k^2]$.
We note that the denominators in $P_0$ and $P_\parallel$ also contains $i\epsilon k^0$.
The second and the third terms are shown not to contribute to the damping rate in Appendix~\ref{app:gauge-inv}, so we omit them from now on.
Thus, the gluon spectral function reduces to 
\begin{align}
\label{eq:gluon-spectrum-1}
\rho^{\mu\nu}_{D}(k)
&= P^{\mu\nu}_\parallel(k_L) \rho_D(k),
\end{align}
where $\rho_D(k)\equiv -2{\text {Im}} [1/(k^2+i\epsilon k^0-\varOmega_{\parallel}(k))]$.
Then, Eq.~(\ref{eq:damping}) becomes
\begin{align}
\label{eq:damping-tracedone}
\begin{split}
\epsilon^L_k \xi_k 
&= g^2\Cf \mf^2 \int_{l} 
 \rho_D(k+l) \rho_S(l_L)\\
&~~~\times \left[R^f(\vk_\perp+\vl_\perp)\right]^2 [\nf(l^0)+\nb(l^0+k^0)]|_{k^0=\eL_k},
\end{split}
\end{align}
where we have evaluated the trace by using $\Tr \left[(\Slash{k}_L+\mf) \gamma^L_\mu(\Slash{l}_L+\mf){\cal P}_+ \gamma^L_\nu \right] P^{\mu\nu}_\parallel(k_L-l_L)= -4\mf^2$, which is obtained by the on-shell conditions $k^2_L=l^2_L=\mf^2$.
We also flipped the sign of $l$ for future convenience.

\subsubsection{$\mf\gg M$ case}

We start with the case that the current quark mass is much larger than the gluon screening mass, $\mf\gg M$.
In this case, the current quark mass regulates all the infrared singularities as will be found later, so we can safely neglect the gluon self-energy:
\begin{align}
\rho_D(k)&= (2\pi)\sgn(k^0)\delta(k^2).
\end{align}
By using this equation and the spectral function of free quark, Eq.~(\ref{eq:free-quark-spectrum}), Eq.~(\ref{eq:damping-tracedone}) becomes
\begin{align}
\label{eq:damping-m>M}
\begin{split} 
\epsilon^L_k \xi_k 
&= g^2\Cf \mf^2 \int \frac{d^4 l}{(2\pi)^4} 
(2\pi)^2\sgn(l^0) \sgn(k^0+l^0) \\
&~~~\times\delta([k+l]^2) \delta(l^2_L-\mf^2)\\
&~~~\times \left[R^f(\vk_\perp+\vl_\perp)\right]^2 [\nf(l^0)+\nb(l^0+k^0)]|_{k^0=\epsilon^L_k} \\
&\simeq \frac{g^2\Cf \mf^2}{8\pi} \int dl^0 
\sgn(l^0) \sgn(\eL_k+l^0)  \sum_{s=\pm 1} \\
&~~~\times\frac{\theta((l^0)^2-\mf^2)\theta\left(\mf^2+\eL_k l^0-sk^3\sqrt{(l^0)^2-\mf^2} \right)}{\sqrt{(l^0)^2-\mf^2}} \\
&~~~\times [\nf(l^0)+\nb(l^0+\eL_k)] ,
\end{split} 
\end{align}
where we have performed the integrations for $\vl^2_\perp$ and $l^3$ by using the two delta functions.
Because the distribution functions give the ultraviolet cutoff at the scale $T$ in $l^0$ integration, $|\vk_\perp+\vl_\perp|^2=2[\mf^2+\epsilon^L_k l^0-sk^3\sqrt{(l^0)^2-\mf^2}] \lesssim T^2\ll eB$ as long as $|k^3|$ is of the order or much smaller than $T$.
Therefore, we have approximated the form factor as unity.
We note that $s=\sgn(l^3)$, so $s$ shows the direction of the movement of the anti-quark whose momentum is $l^3$.
The integration range is shown to be $l^0>\mf$ in Appendix~\ref{app:range}, so we arrive at the expression\footnote{The term containing $\nf$ was already given in Ref.~\cite{Elmfors:1995gr}.
} 
\begin{align}
\label{eq:damping-m>M-2}
\begin{split} 
\epsilon^L_k \xi_k 
&= \frac{g^2\Cf \mf^2}{4\pi} \int^\infty_{\mf} dl^0 
\frac{\nf(l^0)+\nb(l^0+\eL_k)}{\sqrt{(l^0)^2-\mf^2}} .
\end{split} 
\end{align}
The distribution function factor can be rewritten as $\nf(1+\nb)+\nb(1-\nf)$, which shows that the physical process that yields the damping rate above is the pair annihilation of the quark and the anti-quark and its inverse process, which is drawn in the right panel of Fig.~\ref{fig:oneloop-Sigma}.

As we will see later, the dominant contribution to the electrical conductivity comes from the quark whose momentum is of order $T$.
Thus, we focus on the case that $|k^3|\sim T$.
In this case, Eq.~(\ref{eq:damping-m>M-2}) can be evaluated at the leading-log order\footnote{Leading-log approximation means that, we regard $\ln(\epsilon^{-1})$, where $\epsilon$ is a small quantity, as a large number and approximate $\ln(\epsilon^{-1})+{\cal O}(1)\simeq \ln(\epsilon^{-1})$. In the current case, $\epsilon$ is $\mf/T$.
} as
\begin{align}
\label{eq:damping-m>M-final}
\begin{split} 
\epsilon^L_k \xi_k 
&\simeq \frac{g^2\Cf \mf^2}{4\pi} 
\left[\frac{1}{2}+\nb(\eL_k)\right]
\int^T_{\mf} dl^0 
\frac{1}{\sqrt{(l^0)^2-\mf^2}} \\
&\simeq \frac{g^2\Cf \mf^2}{4\pi} 
\left[\frac{1}{2}+\nb(\eL_k)\right]
\ln\left(\frac{T}{\mf}\right),
\end{split} 
\end{align}
 where we have used $\int dl^0/\sqrt{(l^0)^2-\mf^2}=\ln\left(l^0+\sqrt{(l^0)^2-\mf^2}\right)$ and the fact that the dominant contribution comes from the energy region $l^0\ll T$.
 We see that the energy of the quark is $\epsilon^L_k\sim T$ and that of the anti-quark is $l^0\ll T$, which makes the gluon energy $\eL_k+l^0\sim T$.

\subsubsection{$\mf\ll M$ case}

Next, we consider the opposite case, $\mf\ll M$.
In this case, we need to take into account the gluon screening mass:
\begin{align}
\rho_D(k)&= (2\pi)\sgn(k^0)\delta(k^2-M^2),
\end{align}
where 
\begin{align}
\label{eq:gluon-mass}
M^2\equiv \frac{1}{2} \cdot \frac{g^2}{\pi}\sum_f \frac{|\Bf|}{2\pi}.
\end{align}
Here we note that the dispersion relation above has a nonnegligible correction when $k\lesssim \mf$~\cite{Fukushima:2015wck}.
Nevertheless, such low momentum region is found to be irrelevant in the current calculation because the gluon energy is $\eL_k+l^0\gtrsim T$, where we have used $\eL_k\sim T$.
By using this gluon spectral function, Eq.~(\ref{eq:damping-tracedone}) becomes 
\begin{align}
\label{eq:damping-m<M-final}
\begin{split}
\epsilon^L_k \xi_k 
&= g^2\Cf \mf^2 \int\frac{d^4l}{(2\pi)^4} 
(2\pi)\sgn(l^0)\delta(l^2_L-\mf^2)\\
&~~~\times (2\pi)\sgn(k^0+l^0)\delta([k+l]^2-M^2) \\
&~~~\times \left[R^f(\vk_\perp+\vl_\perp)\right]^2 [\nf(l^0)+\nb(l^0+k^0)]|_{k^0=\eL_k} \\
&\simeq \frac{g^2\Cf \mf^2}{4} \int\frac{dl^0 }{(2\pi)} 
\sgn(l^0) \sgn(\eL_k+l^0) \\
&~~~\times \sum_{s=\pm 1} \theta\left(\mf^2+l^0\epsilon^L_k-sk^3\sqrt{(l^0)^2-\mf^2} -\frac{M^2}{2}\right) \\
&~~~\times  \frac{\theta((l^0)^2-\mf^2)}{\sqrt{(l^0)^2-\mf^2}}
 [\nf(l^0)+\nb(l^0+\eL_k)],
\end{split}
\end{align}
where we have approximated $R^f(\vk_\perp+\vl_\perp)\simeq 1$.

Let us consider the case $|k^3|<k_c$ first, where $k_c\equiv M^2\sqrt{A}/(2\mf)\sim M^2/\mf$ with $A\equiv 1-4\mf^2/M^2$.
The integration range is shown to be $l_\pm<l^0$ for $s=\pm\sgn(k^3)$ in Appendix~\ref{app:range}, where $l_\pm$ is defined in Eq.~(\ref{eq:def-lpm}).
Because $l_+\sim M^2T/\mf^2\gg T$, the contribution from $s=\sgn(k^3)$ to Eq.~(\ref{eq:damping-m<M-final}) is exponentially suppressed due to the Fermi/Bose distribution functions.
Thus, Eq.~(\ref{eq:damping-m<M-final}) becomes
\begin{align}
\label{eq:damping-m<M-1}
\begin{split}
\epsilon^L_k \xi_k 
&\simeq \frac{g^2\Cf \mf^2}{4} \int^\infty_{l_-}\frac{dl^0 }{(2\pi)} 
  \frac{\nf(l^0)+\nb(l^0+\eL_k)}{\sqrt{(l^0)^2-\mf^2}} \\
&\simeq \frac{g^2\Cf \mf^2}{8\pi} \left[\frac{1}{2}+\nb(\eL_k)\right]
\ln\frac{T}{l_-+\sqrt{(l_-)^2-\mf^2}} \\
&\simeq \frac{g^2\Cf \mf^2}{4\pi} \left[\frac{1}{2}+\nb(\eL_k)\right]
\ln\frac{T}{M},
\end{split}
\end{align}
where we have used $l_-\simeq M^2[1+(k^3/k_c)^2]/(4|k^3|)\ll T$ for $|k^3|\sim T\gg \mf$, and performed the leading-log approximation in the middle line.
This expression is the same as Eq.~(\ref{eq:damping-m>M-final}) except for the infrared cutoff in the log:
When $\mf\ll M$, the screening mass of the gluon gives the cutoff instead of the current quark mass.

Next we consider the case of $|k^3|>k_c$.
Appendix~\ref{app:range} shows that the integration range is $\mf<l^0<l_-$ and $l_+<l^0$ for $s=-\sgn(k^3)$, and $\mf<l^0$ for $s=\sgn(k^3)$.
Again, the contribution from $l_+<l^0$ can be neglected, and Eq.~(\ref{eq:damping-m<M-final}) becomes
\begin{align}
\label{eq:damping-m<M-2}
\begin{split}
\epsilon^L_k \xi_k 
&\simeq \frac{g^2\Cf \mf^2}{8\pi} \left[\int^{l_-}_{\mf}+\int^{\infty}_{\mf}\right] dl^0  
  \frac{\nf(l^0)+\nb(l^0+\eL_k)}{\sqrt{(l^0)^2-\mf^2}} \\
  &\simeq \frac{g^2\Cf \mf^2}{8\pi}\left[\frac{1}{2}+\nb(\eL_k)\right]
\ln\frac{T}{\mf^2}\left(l_-+\sqrt{(l_-)^2-\mf^2}\right)\\
&\simeq \frac{g^2\Cf \mf^2}{4\pi}\left[\frac{1}{2}+\nb(\eL_k)\right]
\ln\frac{ T}{M} , 
\end{split}
\end{align}
at the leading-log accuracy.
Unexpectedly, the expression is the same as the other case, Eq.~(\ref{eq:damping-m<M-1}).

\section{Results}
\label{sec:results}

Let us evaluate the conductivity by using the expression of the quark damping rate.
First we work in $\mf\gg M$ case.
By using Eq.~(\ref{eq:damping-m>M-final}), Eq.~(\ref{eq:Pi-damping}) yields
\begin{align}
\label{eq:sigma33-result-LL}
\begin{split}
\sigma^{33}
&=  e^2\sum_{f} (q_f)^2 \Nc \frac{|B_f|}{2\pi}
\frac{\beta}{\pi}\\
&~~~\times \int^\infty_{\mf} dk^0 \nf(k^0)[1-\nf(k^0)] 
   \frac{\sqrt{[k^0]^2-\mf^2}}{\xi_k |k^0|} \\
 &= e^2\sum_{f} (q_f)^2 \Nc \frac{|B_f|}{2\pi}
\frac{8\beta}{g^2\Cf \mf^2 \ln\left(T/\mf\right)}\\
&~~~\times \int^\infty_{\mf} dk^0\nf^2(k^0)[1-\nf(k^0)] 
   \frac{\sqrt{[k^0]^2-\mf^2}}{\nb(\eL_k)} \\
&\simeq e^2\sum_{f} (q_f)^2 \Nc \frac{|B_f|}{2\pi}
\frac{4T}{g^2\Cf \mf^2 \ln\left(T/\mf\right)} , 
\end{split}
\end{align}
where we have used the fact that the integrand becomes quickly small at $k^0\ll T$, so that we can safely change the lower bound of the integral to $0$ in the last line.
The integrand is suppressed exponentially also in $k^0\gg T$ region.
Therefore, the dominant contribution of the integral above comes from the energy region $k^0\sim T$.
We also used the formula $\int dk k \nf^2(k)[1-\nf(k)]/\nb(k)=-T^2\nf^2(k)[e^{k/T}(k/T+1)+1]$ in the last line. 

Equation~(\ref{eq:sigma33-result-LL}) is one of the central results of this paper.
Several remarks on this expression are in order:
\begin{enumerate}
\item The conductivity is proportional to $|\Bf|\sim eB$, which is larger than any other scales that have dimension of (energy$)^2$.
It is because of large Landau degeneracy of the quark at LLL in the transverse plane.

\item The $g$-dependence of the conductivity is $\propto g^{-2}$, in contrast to $B=0$ case ($\propto g^{-4}$~\cite{Gagnon:2006hi, Arnold:2000dr}). 
It is due to the fact that the 1 to 2 scattering process instead of 2 to 2 process, which is the leading contribution in $B=0$ case, is responsible for determination of the conductivity.

We can understand how this process is allowed when $B$ is strong as follows:
Let us consider the 1 to 2 process in which a gluon becomes a quark with momentum $k$ and an anti-quark with momentum $l$.
When $B=0$, these particles have dispersion relations in three-dimension, $k^2=l^2=\mf^2$ and $(l+k)^2=0$.
By using the first equation, we see that the last equation can not be satisfied\footnote{Strictly speaking, this equation is satisfied when $\vk$ and $\vl$ are parallel.
However, such configuration would be impossible if the screening effect is taken into account~\cite{Blaizot:1996az}.
}.
Physically, it is clear because a massless particle can not become two massive particles.
By contrast, when we have strong $B$, the quark have one-dimensional dispersion relation while the gluon's dispersion relation is still three-dimensional:
$k^2_L=l^2_L=\mf^2$ and $(k+l)^2_L=(\vk+\vl)^2_\perp$.
We note that these equations are similar to those at $B=0$, except for the right-hand side of the last equation.
These equations can be satisfied, as was seen from the calculation we have done.
It is because the transverse momentum of the gluon, $\vk_\perp+\vl_\perp$, plays a role of gluon mass in the last equation, so that the gluon can decay into two massive particles.

\item The conductivity is proportional to $\mf^{-2}$, so it diverges in the massless limit.
Physically, it is because that the scattering process is forbidden when $\mf=0$~\cite{Smilga:1991xa}:
In the 1 to 2 scattering process drawn in Fig.~\ref{fig:oneloop-Sigma}, the quark and the anti-quark moving in the opposite direction along $B$ in the initial state have the same chirality.
On the other hand, the gluon in the final state has no chirality, so the chirality conservation law forbids such process in the massless limit.
Since $\mf$ is much smaller than other energy scales, this dependence makes the conductivity quite large.

\item The component of the conductivity other than $\sigma^{33}$ vanishes as was discussed after Eq.~(\ref{eq:Wick}).
Physically, it is because that the quarks in the LLL can only move along the direction of the magnetic field, so that the electric charge can not move in the transverse plane.

\item $\ln(T/\mf)$ can be understood from the fact that the ultraviolet cutoff of the energy of the anti-quark coming from the medium is of order $T$ while the infrared cutoff is given by $\mf$, as can be seen from Eq.~(\ref{eq:damping-m>M-final}).

\end{enumerate}

Next we proceed to the case $\mf\ll M$.
Because the quark damping rate is the same as that in $\mf\gg M$ case except for the argument of the log, the conductivity is obtained by changing the infrared cutoff of the log in Eq.~(\ref{eq:sigma33-result-LL}), from $\mf$ to $M$:
\begin{align}
\label{eq:sigma33-result-LL-2}
\begin{split}
\sigma^{33}
&\simeq e^2\sum_{f} (q_f)^2 \Nc \frac{|B_f|}{2\pi}
\frac{4T}{g^2\Cf \mf^2 \ln\left(T/M\right)} .
\end{split}
\end{align}

\section{Ladder diagram summation}
\label{sec:ladder}

\begin{figure*}[t!] 
\begin{center}
\includegraphics[width=0.7\textwidth]{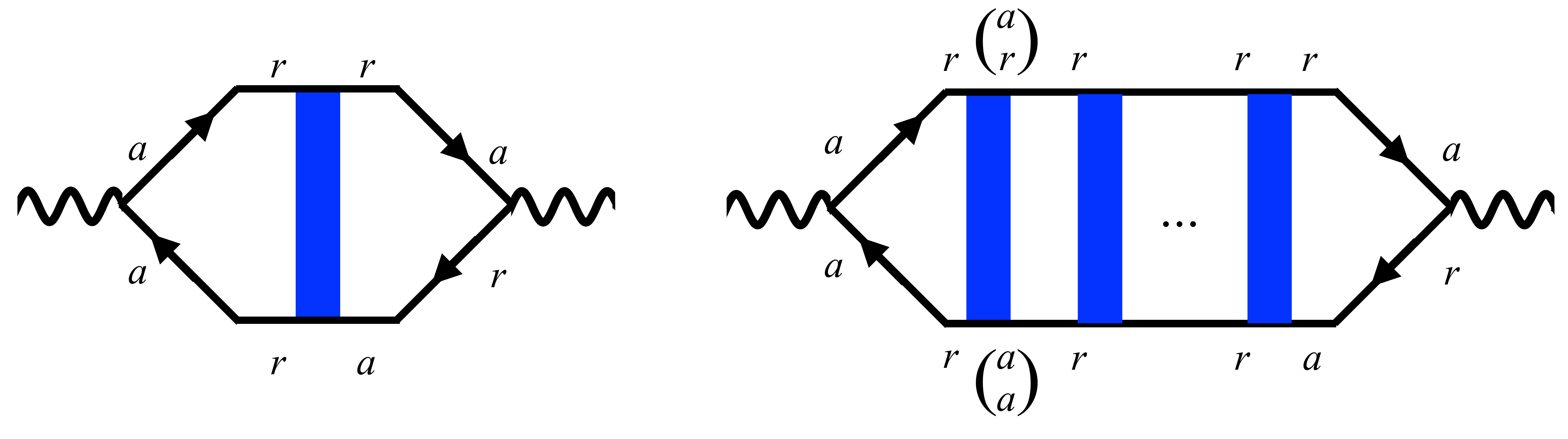} 
\caption{The ladder diagrams with one kernel (left panel) and more kernels (right panel) for $G_{aaar}$. 
The (blue) square is the kernel $K_{ijkl}$ introduced later. 
The two brackets in the right panel represent the two cases indicated in the text.
} 
\label{fig:ladder-Gaaar} 
\end{center} 
\end{figure*} 

\begin{figure}[t!] 
\begin{center}
\includegraphics[width=0.35\textwidth]{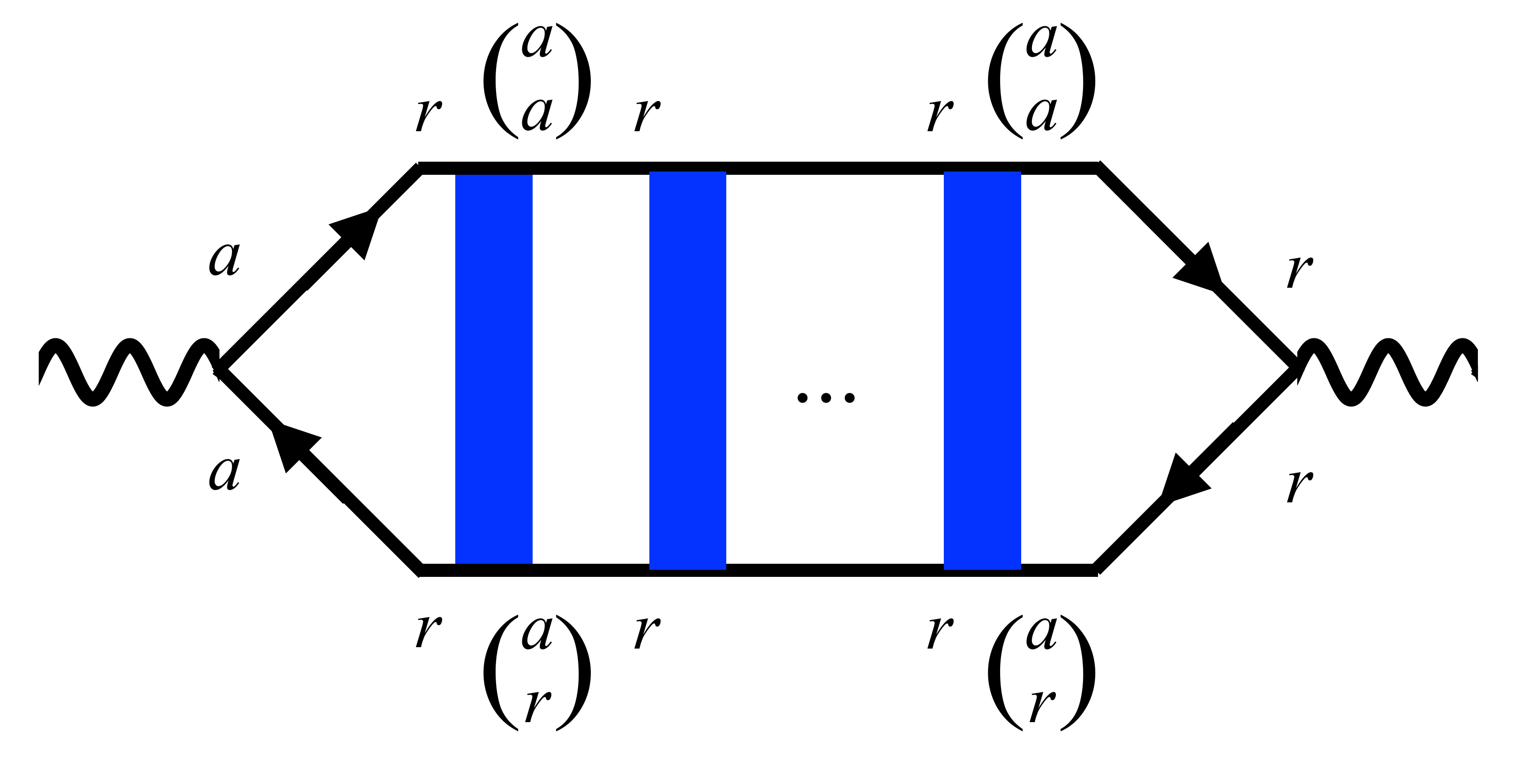} 
\caption{One of the ladder diagram for $G_{aarr}$.
The four brackets represent the two cases indicated in the text.
} 
\label{fig:ladder-Gaarr} 
\end{center} 
\end{figure} 

In the calculation of the electrical conductivity without magnetic field, 
all the ladder diagrams were found to contribute at the leading-order~\cite{Jeon:1994if, Gagnon:2006hi,  Hidaka:2010gh, Wang:2002nba}: 
The suppression of the ladder diagrams by the positive power of $g$ from the kernel 
is cancelled by the negative power of $g$ from the pinch singularity. 
The maximum cancellation happens when the ladder diagram with $n$ kernels contains $(n+1)$ pinch singularities. 
Because of this cancellation, the order of the magnitude of the ladder diagram does not depend on $n$.
This situation persists also in the presence of strong magnetic field, 
so that we need to sum all the ladder diagrams for completing the leading-order calculation.

We start by showing that the terms with $\alpha_2\sim \alpha_7$ and $\beta_2\sim \beta_7$ in Eq.~(\ref{eq:G1122-ra}) do not contribute at the leading order.
Let us consider $G_{aaar}$ as an example. 
As was shown in the previous subsection, this quantity vanishes at one-loop order.
Switching to the ladder diagram, we consider the one with one kernel ($K_{ijkl}$) shown in the left panel of Fig.~\ref{fig:ladder-Gaaar}.
The properties $S_{aa}=0$ and $K_{rrrr}=0$ determine all the indices in this diagram as in the figure.
The two propagators at the left of the kernel has a pinch singularity because there appear $S_{ra}$ and $S_{ar}$ with the same momentum, but the ones at the right does not have the singularity since a pair of $S_{ra}$ appears.
Therefore, the number of the pinch singularity and that of the collision kernel are equal (one in this case).

It is the case also in the diagram with more kernels:
We consider the diagram shown in the right panel of Fig.~\ref{fig:ladder-Gaaar}.
As in the previous diagram, both of the indices in the left part of the leftmost kernel are $r$.
There are two patterns of the configuration of the indices in the right part of the leftmost kernel.
One is the case that both the indices are $a$ ($K_{rraa}$), which makes the left part of the indices of the next kernel $r$.
The other case is that one of the indices is $r$ and the other is $a$.
Also in this case, both the indices of the left part in the next kernel are $r$, in order to have pinch singularity:
From the expression
\begin{align}
\label{eq:Srr}
S^{rr}(k_L) 
&=  \left(\frac{1}{2}-\nf(k^0)\right) 
[S^{ra}(k_L)-S^{ar}(k_L)],
\end{align} 
we see that $S^{rr}$ contains both $S^{ra}$ and $S^{ar}$.
Since the pinch singularity appears from the pair of $S^{ra}$ and $S^{ar}$, the pair of the propagator we are considering should be $S^{rr}$ and $S^{ar}$ because we can not have $S^{ra}$ and $S^{ar}$.
Therefore, the two indices in the left part of the kernel are $r$ in each case.
In the same way, we can apply this argument to the next kernel and find that the indices of the left part in the rightmost kernel are $r$.
Then one can see that the number of the pinch singularity and that of the collision kernel are equal, by using the argument used in the case of one kernel.
The situation is the same also for the other terms than $G_{aaar}$.

Now let us discuss for $G_{aarr}$, by taking a look at the diagram drawn in Fig.~\ref{fig:ladder-Gaarr}.
In the same way as in $G_{aaar}$, the leftmost kernel is $K_{rraa}$ or $K_{rrar}$, and to have pinch singularity, the indices of the left part of the next kernel are $(rr)$.
By repeating the same argument, we can list all the configurations of the indices as in the figure.
In any of these configurations, we see that there are $(n+1)$ pinch singularities for the ladder diagram with $n$ kernels.
Thus, we see that $G_{aarr}$ is much larger than $G_{aaar}$ and the other terms in Eq.~(\ref{eq:G1122-ra}).

\begin{figure}[t!] 
\begin{center}
\includegraphics[width=0.4\textwidth]{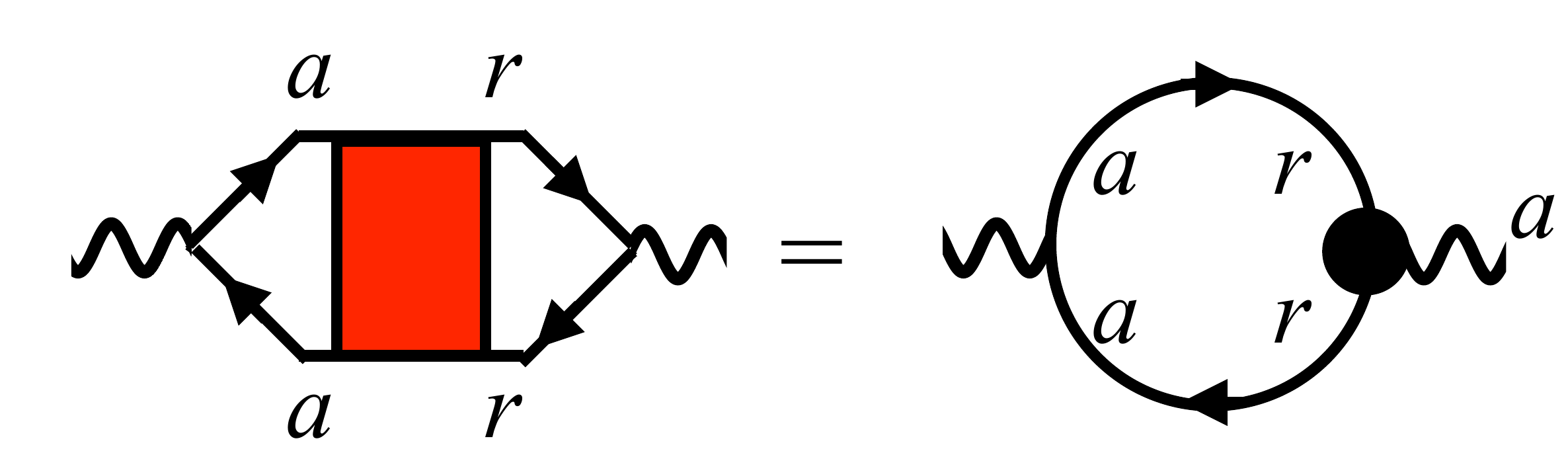} 
\caption{The relation between $G_{aarr}$ and $\varGamma$.
The (red) square represents $G_{aarr}$.
The black blob represents a vertex function $\varGamma^\mu$.
} 
\label{fig:G-Gamma} 
\end{center} 
\end{figure}

\begin{figure}[t!] 
\begin{center}
\includegraphics[width=0.45\textwidth]{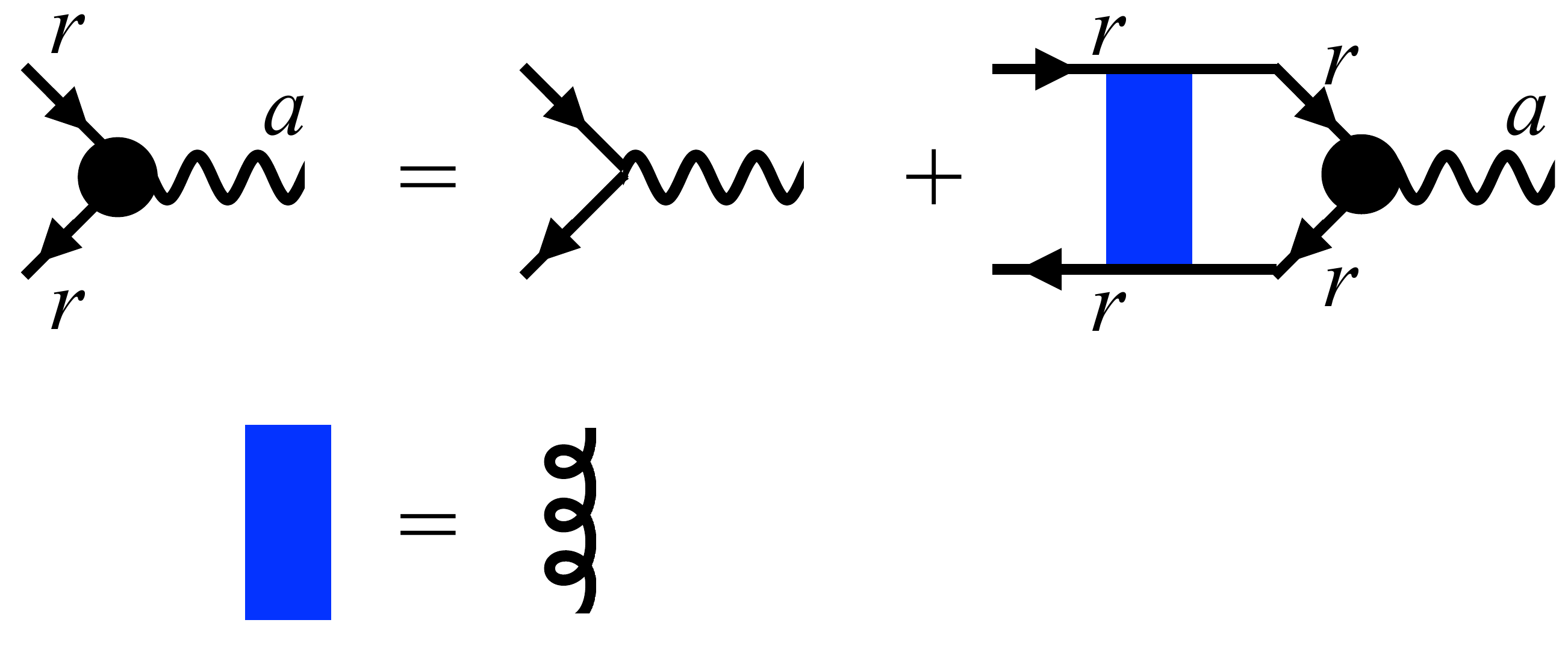} 
\caption{Bethe-Salpeter equation (upper panel) and the explicit form of the kernel for the one gluon exchange (lower panel).
} 
\label{fig:BSeq-vertex} 
\end{center} 
\end{figure}

Now we can proceed the calculation.
For the convenience to check the Ward-Takahashi identity, we introduce the quark-photon vertex function $\varGamma^\mu(k)$, which is related to the four-point function as 
\begin{align}
\int_{l^2,l_L}G^{\mu\nu f}_{aarr}(k,k,l,l)
&= \Tr\left[\gamma^\mu S^A(k_L) \varGamma^\nu(k)S^R(k_L)\right],
\end{align}
which is represented diagrammatically in Fig.~\ref{fig:G-Gamma}.
From this figure, one sees that this vertex function has the indices $(rra)$ in r/a basis.
The integral equation (Bethe-Salpeter equation) that is used to sum all the ladder diagrams, whose diagrammatic expression is in Fig.~\ref{fig:BSeq-vertex}, reads
\begin{align}
\varGamma^\mu(k)
&= \gamma^\mu 
+\int_l \sum_{\alpha,\beta=r,a} 
[K_{rr\alpha\beta}(k,l)S^{\alpha r}(l_L)\varGamma^\mu(l)S^{r\beta}(l_L)].
\end{align}
We note that the kernel $K_{rr\alpha\beta}$ generally has a spinor structure.
Also, notice that only the vertex with the indices $(rra)$ appears:
One may think that the other nonzero vertex, which has the indices $(aaa)$, can appear in the second term in the right-hand side.
Nevertheless, in that case, the property $S^{aa}=0$ forces us to make the indices of the kernel as $K_{rrrr}$, which vanishes identically, so only the vertex with the indices $(rra)$ appear at the equation above.
By using Eq.~(\ref{eq:Srr}), we pick up only the pairs of $S_{ar}(l_L)$ and $S_{ra}(l_L)$, which generate pinch singularity, in the summation of $\alpha, \beta$ above:
\begin{align}
\label{eq:BS-1}
\begin{split}
\varGamma^\mu(k)
&= \gamma^\mu 
+\int_l \Bigl[\Bigl\{K_{rraa}(k,l) \\
&~~~+\left(\frac{1}{2}-\nf(l^0)\right)\left(K_{rrar}(k,l)-K_{rrra}(k,l)\right) \Bigr\} \\
&~~~\times S^{a r}(l_L)\varGamma^\mu(l)S^{ra}(l_L)\Bigr].
\end{split}
\end{align}

The kernel is given by the one-gluon exchange process at the leading order (lower panel of Fig.~\ref{fig:BSeq-vertex}).
With this kernel, Eq.~(\ref{eq:BS-1}) becomes
\begin{align}
\label{eq:BS-2}
\begin{split}
\varGamma^\mu(k)
&= \gamma^\mu 
-g^2\Cf\int_l [R^f(\vk_\perp-\vl_\perp)]^2 
\Bigl\{ D^{\alpha\beta}_{rr}(k-l) \\
&~~~ +\left(\frac{1}{2}-\nf(l^0)\right)\left(D^{\alpha\beta}_{ra}(k-l)-D^{\alpha\beta}_{ar}(k-l)\right) \Bigr\} \\
&~~~\times \gamma_\alpha S^{a r}(l_L)\varGamma^\mu(l)S^{ra}(l_L) \gamma_\beta \\ 
&= \gamma^\mu 
+g^2\Cf\int_l   \gamma_\alpha S^{A}(l_L)\varGamma^\mu(l)S^{R}(l_L) \gamma_\beta  \rho^{\alpha\beta}_{D}(k-l) \\
&~~~ \times\left(\nf(-l^0)+\nb(k^0-l^0)\right) [R^f(\vk_\perp-\vl_\perp)]^2 ,
\end{split}
\end{align}
where we have used $D^{\mu\nu}_{rr}(k)=-i[1/2+\nb(k^0)][D^{\mu\nu}_{R}(k)-D^{\mu\nu}_{A}(k)]$ and $D^{\mu\nu}_{ra/ar}=-iD^{\mu\nu}_{R/A}$ in the last line.
This equation is shown to be consistent with the Ward-Takahashi (WT) identity in Appendix~\ref{app:WT}.
We note that only the on-shell gluon appears at the integral above.
We decompose the spinor structure as 
$S^A(k_L)\varGamma^\mu(k)S^R(k_L)=(\Slash{k}_L+\mf){\cal P}_+\rho_S(k_L)A^\mu(k)/2$, 
where $A^\mu$ does not have a spinor structure. 
Then, the integral equation for $A^\mu$ becomes
\begin{align}
\label{eq:BS-3}
\begin{split}
\xi_k k^0 A^\mu(k)
&=k^\mu
-g^2\Cf\mf^2\int_l \rho_S(l_L)\rho_D(k-l) A^\mu(l) \\
&~~~\times\left(\nf(-l^0)+\nb(k^0-l^0)\right) [R^f(\vk_\perp-\vl_\perp)]^2,
\end{split}
\end{align}
where we have used Eqs.~(\ref{eq:S-spinorstr}), (\ref{eq:gluon-spectrum-1}), $(\Slash{k}_L+\mf)\gamma^\mu(\Slash{k}_L+\mf)=2k^\mu(\Slash{k}_L+\mf)$, and $\gamma_\alpha (\Slash{l}_L+\mf)\gamma_\beta P^{\alpha\beta}_{\parallel}(k-l)=-(\Slash{k}_L+\mf)$, which are valid for the on-shell  $k_L$ and $l_L$.
From the expression above, we see that 
\begin{align}
A^3(k^0,k^3,\vk_\perp)
&=-A^3(k^0,-k^3,\vk_\perp)
=-A^3(-k^0,k^3,\vk_\perp),
\end{align}
by looking at the integral kernel.
By changing the sign of $l$, and using this property and the expression of the damping rate (\ref{eq:damping-tracedone}), Eq.~(\ref{eq:BS-3}) for $\mu=3$ becomes
\begin{align}
\label{eq:BS-4}
\begin{split}
k^3
&=g^2\Cf\mf^2\int_l \rho_S(l_L)\rho_D(k+l) [A^3(k)-A^3(l)] \\
&~~~\times\left(\nf(l^0)+\nb(k^0+l^0)\right) [R^f(\vk_\perp+\vl_\perp)]^2.
\end{split} 
\end{align}
This is the integral equation that is necessary for the complete leading-order analysis.
From the definition of the $ A^\mu(k)$ below Eq.~(\ref{eq:BS-2}), 
the relation to the four-point function is
\begin{align}
\label{eq:Gaarr-A}
\int_{l^2,l_L}G^{\mu\nu f}_{aarr}(k,k,l,l)
&= \rho_S(k_L)k^\mu A^\nu(k),
\end{align}
One can see that these equations are equivalent to the linearized Boltzmann equation in Appendix~\ref{app:linear-Boltzmann}.

Now we perform iteration in Eq.~(\ref{eq:BS-4}), to see the effect of the ladder diagram summation.
In $\mf\gg M$ case, from Eq.~(\ref{eq:damping-m>M-final}), the zeroth order solution reads
\begin{align}
\label{eq:A-zero}
\begin{split}
A^3_{(0)}(k)&=\frac{k^3}{\xi_k k^0}\\
&=  \frac{8\pi k^3\nf(k^0)}{
g^2\Cf \mf^2 \nb(k^0)
\ln\left(T/\mf\right)} . 
\end{split}
\end{align}
By iterating it into Eq.~(\ref{eq:BS-4}) we have
\begin{align}
\begin{split}
&g^2\Cf\mf^2\int_l \rho_S(l_L)\rho_D(k+l) A^3_{(0)}(l) \\
&~~~\times\left(\nf(l^0)+\nb(k^0+l^0)\right) [R^f(\vk_\perp+\vl_\perp)]^2\\
&= \frac{1}{\ln\left(T/\mf\right)}\int^\infty_{\mf} dl^0 
\sum_{s=\pm 1} s \\
&~~~\times\left(\nf(l^0)+\nb(k^0+l^0)\right) \frac{\nf(l^0)}{\nb(l^0)},
\end{split} 
\end{align}
which is found to be zero after summing over $s$, corresponding to the cancellation of the contributions coming from positive and negative $l^3$.
Therefore, the ladder summation is not necessary at the leading order calculation.

On the other hand, in $\mf \ll M$ case, the zeroth order solution is $A^3_{(0)}(k)= 8\pi k^3\nf(\eL_k)/[g^2\Cf \mf^2 \nb(\eL_k)\ln\left(T/M\right)] $, and the first iteration reads
\begin{align}
\label{eq:m<<M-iteration}
\begin{split}
&g^2\Cf\mf^2\int_l \rho_S(l_L)\rho_D(k+l) A^3_{(0)}(l) \\
&~~~\times\left(\nf(l^0)+\nb(k^0+l^0)\right) [R^f(\vk_\perp+\vl_\perp)]^2\\
&= \frac{1}{\ln\left(T/M\right)}
 \int^\infty_{l_-} dl^0
 \frac{ \nf(l^0)}{\nb(l^0)} 
\left(\nf(l^0)+\nb(k^0+l^0)\right), 
\end{split} 
\end{align}
where we have considered the case $|k^3|<k_c$ and retained only $s=\sgn(k^3)$ term.
The integral above does not contain any infrared singularity, so this quantity is of order $T/\ln(T/M)$.
This is smaller than the left-hand side of Eq.~(\ref{eq:BS-4}), $k^3\sim T$, by the factor of $[\ln(T/M)]^{-1}$, so taking only the zeroth order solution (\ref{eq:A-zero}) is enough at the leading-log accuracy.
One can show that the ladder summation is unnecessary also in $|k^3|>k_c$ case in the same way.

Finally, we verify the assumption made above Eq.~(\ref{eq:Pimunu-A}), 
that is, the integral $\int_{l^2}G^{\mu\nu f}_{1122}(k,k,l,l) $ does not depend on $k^2$. 
According to Eq.~(\ref{eq:Gaarr-A}), 
this verification is accomplished if one could show that $A^\nu(k)$ does not depend on $k^2$. 
In both of $\mf\gg M$ and $\mf\ll M$ cases, the zeroth-order solution $A^3_{(0)}(k)$ does not depend on $k^2$.
In the former case, this completes the confirmation because $A^3_{(0)}(k)$ is already the full solution at the leading order.
In the latter case, we see that iteration does not yield $k^2$ dependence by looking at Eq.~(\ref{eq:m<<M-iteration}), so the assumption is valid also in this case.

\section{Implication in heavy ion collision}
\label{sec:heavyion}

In this section, we briefly discuss possible implications 
for the heavy ion collisions from our results obtained in the previous section. 
We evaluate the value of the conductivity by using typical values of $T$, $B$, $g$, and $\mf$ realized in the heavy ion collision, 
and make a comparison with the preceding results at $B=0$.
We also discuss the back reaction from the induced current to the electromagnetic field.
Furthermore, we discuss the effect of magnetic field on the production rate of soft dileptons.

\subsection{Evaluation of conductivity}

Having the heavy ion collisions in mind, we use the following values for the parameters: 
\begin{align}
\label{eq:parameters}
\begin{split}
\alpha_s&\equiv\frac{g^2}{4\pi}=0.3,\\
\mf&= 3 \ {\text {MeV ($u$)} }, \ 5 \ {\text {MeV ($d$)}}, ~100 \ {\text {MeV ($s$)}}
\\
eB&= 10m^2_\pi= (443 \ {\text {MeV}})^2,
\end{split}
\end{align}
where $m_\pi=140$ MeV is the pion mass.
In order to highlight the effect of the magnetic field, we have taken the very large value for $eB$. 
With these parameters, the gluon screening mass (\ref{eq:gluon-mass}) is approximately $M\simeq 140$ (160) MeV in $\Nf=2$ (3) case.
Because it satisfies $\mf<M$, we use our results (\ref{eq:sigma33-result-LL-2}) in the case $\mf\ll M$ from now on.

First, let us compare the parametric behaviors of our result at finite $ B$ and of the result at $B=0$.
From Eq.~(\ref{eq:sigma33-result-LL-2}), the conductivity along the magnetic field is parametrically evaluated as $\sigma^{33}/e^2\sim eB T/[g^2\mf^2\ln (T/M)]$.
On the other hand, the parametric estimate for $B=0$ is $\sigma_{B=0}/e^2\sim g^{-4}T(\ln g^{-1})^{-1}$~\cite{Gagnon:2006hi, Arnold:2000dr}.
Thus their ratio is given by 
\begin{align}
\sigma^{33}/\sigma_{B=0}
\sim g^2 \frac{eB}{\mf^2} 
\simeq 8.2\times 10^4 ,
\end{align} 
where we have neglected the log factor and used the parameters given in Eq.~(\ref{eq:parameters}).
We see that the conductivity in the strong magnetic field is much larger than that without the magnetic field, 
mainly because of the small value of the current quark mass.

\begin{figure}[t!] 
\begin{center}
\includegraphics[width=0.48\textwidth]{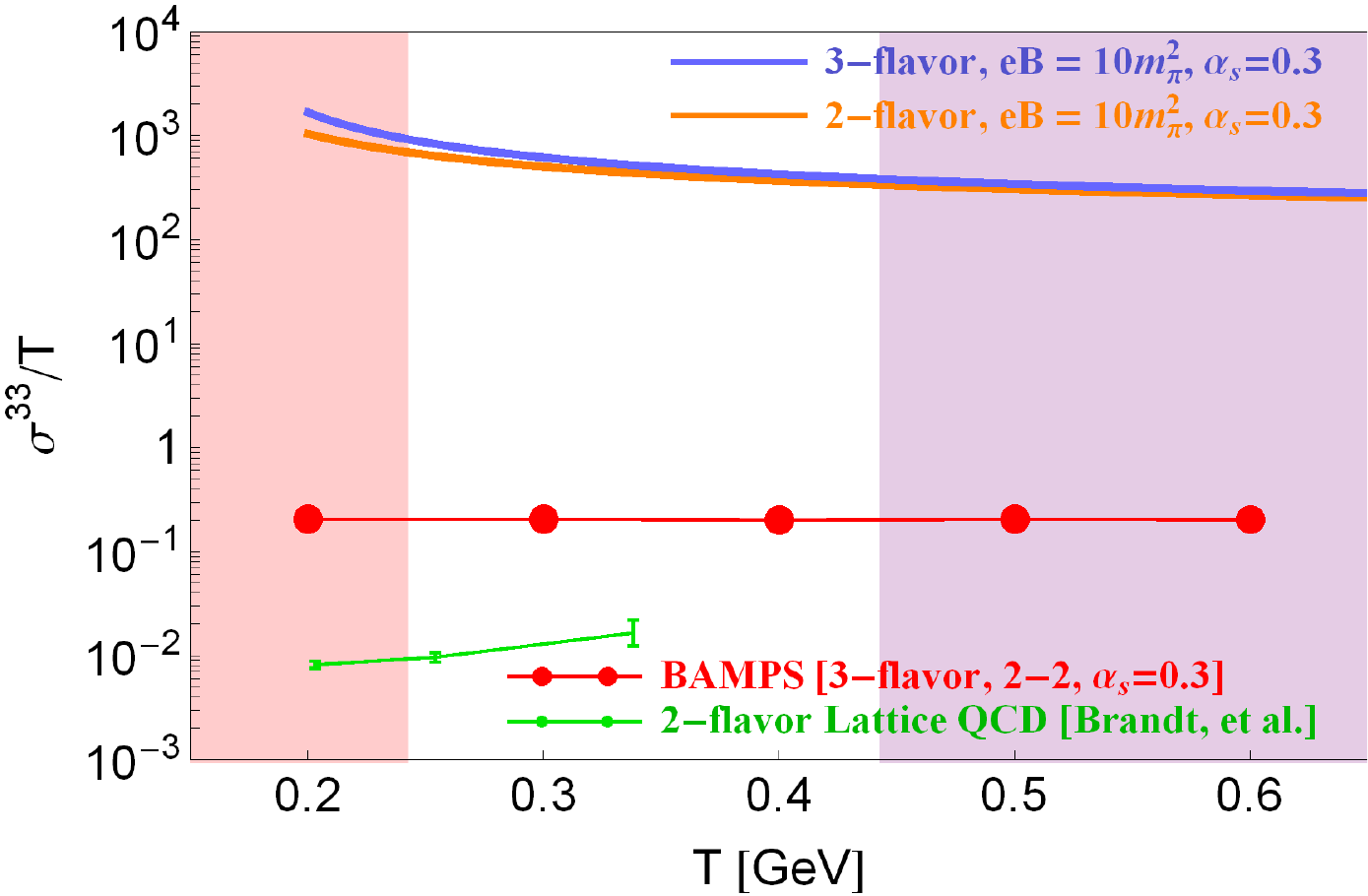} 
\caption{
The orange and blue curves shows our results (\ref{eq:sigma33-result-LL-2}) for two- and three-flavor cases, respectively. 
The parameters, $g$, $\mf$, and $eB$, are set to the values given in Eq.~(\ref{eq:parameters}). 
The red (purple) area in the left (right) part shows the temperature range 
that does not satisfy $\sqrt{\alpha_s eB}<T$ ($T<\sqrt{eB}$).
The red points are the result of BAMPS at $B=0$ for the massless three-flavor case~\cite{Greif:2014oia}, 
in which $\alpha_s$ is fixed with the value in Eq.~(\ref{eq:parameters}). 
The green points with statistical error bars are the result of two-flavor lattice QCD at $B=0$ 
with pion mass $ \sim 270$ MeV~\cite{Brandt:2015aqk}.
} 
\label{fig:sigma_vs_g} 
\end{center} 
\end{figure} 


To show our results in a more useful form for the phenomenology 
and to explicitly indicate the region of validity of our calculation, 
we show a plot of $\sigma^{33}$ as a function of $T$ 
for two- and three-flavor cases ($\Nf=2$ and $3$) in Fig.~\ref{fig:sigma_vs_g}. 
In Eq.~(\ref{eq:sigma33-result-LL-2}),  we take $\Nc=3$ and $e^2=0.092$ from $\alpha\equiv e^2/(4\pi)=1/137$. 
The values for $g$, $\mf$, and $\sqrt{eB}$ are given in Eq.~(\ref{eq:parameters}). 
The red and purple areas are shown to exclude the temperature ranges 
in $T<\sqrt{\alpha_s eB}$ and $\sqrt{eB}<T$, respectively, 
where our assumption $M\ll T\ll \sqrt{eB}$ is not valid. 
Therefore, our result is expected to be reliable only in the window in between, and 
there are crossovers to the regions of other hierarchies near the boundaries. 
The reference results at $B=0$ are taken 
from the numerical solution of the Boltzmann equation\footnote{We could not find the numerical prefactor 
of the conductivity at $B=0$ in perturbative QCD, 
so instead we plot the value obtained with this model.} (BAMPS)~\cite{Greif:2014oia} 
for massless three-flavor quarks with 2 to 2 collision effects 
and from the two-flavor lattice QCD simulation \cite{Brandt:2015aqk} as an example of nonperturbative calculations. 
Though there is a deviation between these two reference results by a factor of $\sim 10$, 
both of them are almost constant in all the temperature range, 
which is consistent with the parametric behavior informed from the perturbative calculations discussed above. 
Compared with the result of BAMPS (lattice) at $B=0$, 
our result is larger by a factor $\sim 10^4$ ($\sim 10^5$) in all the temperature range. 
Thus again, we see that the strong magnetic field significantly increases the conductivity. 
Nevertheless, when $T$ is as large as or much larger than $\sqrt{eB}$ (the purple area in the figure), 
the higher LL, whose scattering process is not suppressed by $\mf^2$, is expected to contribute to the conductivity.
Therefore, $\sigma^{33}$ in such temperature region is likely to be much smaller than our result, 
so that the smooth crossover from our result to the perturbative result at $B=0$ (BAMPS) is realized.

We also comment on the contribution from $s$ quark:
Our result~(\ref{eq:sigma33-result-LL-2}) is proportional to $\mf^{-2}$, 
which is therefore quite sensitive to the current quark mass when it is small. 
Because of this enhancement, the contributions of the $u$ and $d$ quarks dominate over the $ s$-quark contribution. 
In Fig.~\ref{fig:sigma_vs_g}, we confirm that the two-flavor result is barely changed by adding the $ s$-quark contribution.

\subsection{Back reaction from induced current to electromagnetic field}
We briefly discuss the effect of the induced current to the dynamics of the electromagnetic field.
The Maxwell equations read
\begin{align}
\nabla\times \vE&= -\partial_t \vB,\\
\partial_t E^i&= (\nabla\times \vB)^i-\sigma^{ij} E^j.
\end{align}
When $\sigma^{ij}=\sigma \delta^{ij}$, which is the case when $E$ and $B$ are weak, these equations lead to $\nabla^2\vB=(\partial^2_t+\sigma\partial_t)\vB$.
If $\sigma$ is large enough, it reduces to 
\begin{align}
\nabla^2\vB=\sigma\partial_t\vB.
\end{align}
This expression indicates that the lifetime of the magnetic field is parametrically $\tau \sim\sigma L^2$, where $L$ is the characteristic length of the system~\cite{McLerran:2013hla, Tuchin:2013apa, Tuchin:2015oka}.

On the other hand, when $\sigma^{ij}=\sigma^{33} \delta^{i3}\delta^{j3}$, which is realized when $B$ is quite strong, we get the following equation for $B^3$ from the Maxwell equations:
\begin{align}
\nabla^2 B^3=\partial^2_t B^3.
\end{align}
This equation does not contain $\sigma^{33}$, 
so the back reaction to the 3-component of the magnetic field is found to be absent in the LLL approximation. 
For the current to have the transverse components, 
the transitions between the Landau levels need to be activated. 
Therefore, more quantitatively, the other components of $\sigma^{ij}$ is suppressed 
by the Boltzmann factor $e^{-\sqrt{eB}/T}$, so also the back reaction 
from the induced current to $B^3$ should be suppressed by the same factor.
This suppression persists until the magnetic field becomes weaker 
and becomes of order $\sqrt{eB}\sim T$ in the time evolution of the heavy ion collision.

\subsection{Soft dilepton production rate}

It was discussed that the conductivity and the production rate of soft dilepton are related in the case of $B=0$~\cite{Moore:2006qn}.
We apply the same argument to our case, $B\neq 0$.
The production rate of the dilepton with momenta $p_1$ and $p_2$ reads
\begin{align}
\frac{d\varGamma}{d^4p}
&= -\frac{\alpha}{24\pi^4 p^2}
g_{\mu\nu} \varPi^{\mu\nu}_{12}(p),
\end{align}
where  $p\equiv p_1+p_2$ is a momentum of the virtual photon that decays into the dilepton.
When $p$ is large enough so that the effect of $B$ on the virtual photon is negligible, we expect that this expression is reliable.
This condition is parametrically, $p\gg e\sqrt{eB}$, 
which is the energy scale of the photon self-energy due to the magnetic field~\cite{Fukushima:2011nu}.

The current correlator has a form $\varPi^{\mu\nu}_{12}(p)=\varPi^{\parallel}_{12}(p)P^{\mu\nu}_{\parallel}(p)$ for general $p$ at the LLL approximation~\cite{Fukushima:2011nu, Hattori:2012je}.
Because we have Eq.~(\ref{eq:sigma-Pi12}) in $|\vp|=0$ limit, we have 
\begin{align}
\varPi^{\parallel}_{12}(\omega,\vp=\vzero)
&= 2T\sigma^{33}.
\end{align}
We also need to assume $p\ll \xi_k\sim g^2\mf^2/T\ln(T/M)$ to apply this result, in which the collision effect is essentially important.

Summarizing these expressions, 
we obtain the result for the dilepton production rate at $\vp =\vzero$ as 
\begin{align}
\frac{d\varGamma}{d^4 p}
&= \frac{\alpha}{12\pi^4 \omega^2}
T\sigma^{33},
\end{align}
for $\omega$ satisfying $e\sqrt{eB} \ll \omega\ll  g^2\mf^2/T\ln(T/M)$. 
Here we have used $g_{\mu\nu}P^{\mu\nu}_\parallel(p)=-1$.
This expression shows that the production rate is proportional to the conductivity, which is a large value, so it suggests that the production of the soft dilepton may be significantly enhanced by the magnetic field.

We note that there are difference of factor 3 compared with the expression for $B=0$ case~\cite{Moore:2006qn}.
It is because, the conductivity tensor is isotropic when $B=0$ so that there are three nonzero components ($x, y,$ and $z$), while it has only one nonzero component ($z$), whose direction is along the magnetic field, in the presence of strong magnetic field.

\section{Summary and Concluding Remarks}
\label{sec:summary}

We computed the electrical conductivity of QGP in magnetic field by using LLL approximation, starting from quantum field theory by taking into account 1 to 2 scattering process for $\mf\gg M$ and $\mf\ll M$ cases.
We showed that the one-loop approximation suffices 
at the full leading order for $\mf \gg M$ case, and  
at the leading-log approximation for $\mf \ll M$ case. 
We found that the conductivity tensor is nonzero only in $(33)$ component, and it is quite large value mainly due to small current quark mass.
We also discussed possible implications to the heavy ion collision experiment, such as the back reaction of the induced current to the electromagnetic field and the soft dilepton production rate.

Our result suggests that $\sigma^{33}$ is also enhanced by 
the large degeneracy-factor of $eB/(2\pi)$, when $B$ is strong enough.
This behavior is in contrast with the results from the lattice QCD~\cite{Buividovich:2010tn} and the Boltzmann equation~\cite{Harutyunyan:2016rxm, Kerbikov:2014ofa} for weak $B$, which suggest that the conductivity is independent of $B$.

Here a few remarks on implications of our result are in order: 
 As was mentioned in the Introduction, the electrical conductivity is an important quantity in the magnetohydrodynamics.
 Implementation of theoretical prediction of this quantity, including ours, needs to be done in the numerical simulation.
Also, it was suggested that the anisotropy of the conductivity tensor yields the elliptic flow $(v_2)$ of the photon~\cite{Yin:2013kya}.
Our result shows very strong anisotropy, so it may have large effect on the photon $v_2$.
Another application is directed to the Dirac semimetal realized in condensed matter experiment~\cite{Li:2014bha}. 
The quasiparticles appearing in this material has properties of a chiral fermion with the relativistic dispersion relation.  
The energy scale of the magnetic fields applied in the experiments is much larger than the temperature:
$T=20$ K and $B=2\sim9$ T are realized in Ref.~\cite{Li:2014bha}, so the energy scales for the temperature and the magnetic field are $k_B T=1.7\times 10^{-3}$ eV and $\sqrt{eB\hbar c^2}=11\sim 23$ eV, respectively.
Thus, our assumption ($k_B T\ll \sqrt{eB\hbar c^2}$) is expected to be satisfied, 
and our formalism may be applicable to this system, 
though the explicit expression for the quark damping rate may need to be modified depending on 
the specific form of interactions. 

We have not gone beyond the leading-log approximation, for which one needs to fully evaluate the quark damping rate and solve the integral equation (\ref{eq:BS-4}) for $\mf\ll M$ case.
We also need to consider 2 to 2 scattering effect at this order\footnote{
We thank Ho-Ung Yee for pointing out this fact.}.
Also, to explore the intermediate regime  $\sqrt{eB} \sim T$, one needs to go beyond the LLL approximation. 
Finally, analyzing the back reaction from the induced current to the electric field and the transverse components of the magnetic field ($B^1, B^2$) would be an interesting task. 
We leave these interesting tasks to a future work.

\section*{ACKNOWLEDGMENTS}

D.S. thanks Dirk Rischke and Shi Pu for fruitful discussion.
We thank Moritz Greif for providing us with the numerical data for the conductivity evaluated with BAMPS.
D.S. is supported by the Alexander von Humboldt Foundation.
K.H. is supported by China Postdoctoral Science Foundation under Grant No.~2016M590312 
and, at the early stage of this work, by Japan Society for the Promotion of Science Grants-in-Aid No.~25287066. 
K.H. is also grateful for support from RIKEN-BNL Research Center. 

\appendix

\section{Gauge fixing paremeter invariance}
\label{app:gauge-inv}

The gluon spectral function generated from Eq.~(\ref{eq:gluon-propagator}) is,
\begin{align}
\begin{split}
\rho^{\mu\nu}_D(k)
&= 2\pi\sgn(k^0)k^\mu k^\nu 
\Biggl[\delta(k^2_L)\left\{\frac{1}{\vk^2_\perp}+{\text{Re}}\frac{1}{k^2-\varOmega_\parallel(k)}\right\}\\
&~~~-\frac{\delta(k^2)}{\vk^2_\perp}
+(\alpha-1)\delta'(k^2)
\Biggr]
+{\text{[Eq.~(\ref{eq:gluon-spectrum-1})}]},
\end{split}
\end{align}
where we have used Im$[(k^0+i\epsilon)^2-\vk^2]^{-2}=\pi\sgn(k^0) \delta'(k^2)$.
We note that they are generated from the gauge fixing term and the denominator of the projection operators in Eq.~(\ref{eq:gluon-propagator}).
We see that all the terms are proportional to $k^\mu k^\nu$ as long as $\mu,\nu=0, 3$, which is the case in the calculation of the quark damping rate in the LLL approximation.
Therefore, the trace in Eq.~(\ref{eq:damping}) becomes proportional to 
\begin{align}
\begin{split}
&\Tr \left[(\Slash{k}_L+\mf)
\gamma^L_\mu(\Slash{l}_L+\mf){\cal P}_+ \gamma^L_\nu \right] (k-l)^\mu (k-l)^\nu \\
&= \frac{1}{2}\Tr \left[ \Slash{k}_L
\gamma^L_\mu \Slash{l}_L \gamma^L_\nu
+\mf^2 \gamma^L_\mu\gamma^L_\nu \right] (k-l)^\mu (k-l)^\nu \\
&= 2\{2[k_L\cdot(k-l)_L][l_L\cdot(k-l)_L] \\
&~~~-(k-l)^2_L[k_L\cdot l_L-\mf^2] \},
\end{split}
\end{align}
which vanishes by using the on-shell conditions for $k_L$ and $l_L$.
Therefore, only Eq.~(\ref{eq:gluon-spectrum-1}) contributes to the quark damping rate, and the other terms in the gluon spectral function do not.

The contribution to the ladder summation (Eq.~(\ref{eq:BS-2})) is also found to be zero in the same way.

\section{Integration range in 1 to 2 scattering process}
\label{app:range}

We evaluate the range of $l^0$ integral in the collision kernel of 1 to 2 scattering process in this Appendix.

\subsection{$\mf\gg M$ case}

We start with the $\mf\gg M$ case.
The two step functions in Eq.~(\ref{eq:damping-m>M}) determines the integration range.
The first one gives
\begin{align}
\label{eq:l0-range-1}
l^0<\mf, ~~\mf<l^0.
\end{align}
The second one leads to 
\begin{align}
f(l^0)
\equiv \mf^2+\eL_k l^0-sk^3\sqrt{(l^0)^2-\mf^2}>0,
\end{align}
where $s$ is the sign of $l^3$.
The solution of $f=0$ is $l^0=-\eL_k$ when $s=-\sgn(k^3)$, and does not exist when $s=\sgn(k^3)$.
It shows that $f(l^0)$ does not cross the $y$ axis at positive $l^0$, so the properties $f(\mf)=\mf(\mf+\eL_k)>0$ and $f(\infty)=\infty$ lead us to the result that $f(l^0)$ is larger than zero for positive $l^0$.
For negative $l^0$, we have the properties $f(-\mf)=-\mf(\eL_k-\mf)<0$ and $f(-\infty)=-\infty$.
At most only one solution of $f=0$ exists in the negative $l^0$ region, so the properties above show that $f(l^0)<0$ for negative $l^0$.
Summarizing these observation and considering Eq.~(\ref{eq:l0-range-1}), we see that the integration range is $l^0>\mf$.

\subsection{$\mf\ll M$ case}
Next, we evaluate the integration range in the case of $\mf\ll M$.
In this case, the function $f$ is modified by the effect of $M$ as
\begin{align}
f(l^0)
&\equiv \mf^2+\eL_k l^0-sk^3\sqrt{(l^0)^2-\mf^2}-\frac{M^2}{2}.
\end{align}
Its value at a few specific points is $f(\pm \infty )=\pm \infty$ and $f(\pm \mf)=\mf^2\pm\mf\eL_k -M^2/2$.
$f(-\mf)$ is always negative because of $M\gg \mf$, and $f(\mf)$ is negative (positive) when $|k^3|<k_c$ ($|k^3|>k_c$), where $k_c\equiv M^2\sqrt{A}/(2\mf)$ with $A\equiv 1-4\mf^2/M^2$.

\subsubsection{$|k^3|<k_c$ case}
When $|k^3|<k_c$, the solution of $f(l^0)=0$ is $l^0=l_\pm$ for $s=\pm \sgn(k^3)$, where 
\begin{align}
\label{eq:def-lpm}
l_\pm\equiv \frac{M^2}{2\mf^2}
\left[\eL_k\left(1-\frac{2\mf^2}{M^2}\right)\pm|k^3|\sqrt{A}\right].
\end{align}
Combining this property and the behaviors above, we see that the range where $f(l^0)>0$ is satisfied is
\begin{align} 
l_\pm<l^0,
\end{align}
for $s=\pm \sgn(k^3)$.

\subsubsection{$|k^3|>k_c$ case}
When $|k^3|>k_c$, the solution of $f(l^0)=0$ is $l^0=l_+$ and $l^0=l_-$ for $s= \sgn(k^3)$, and there is no solution for $s=- \sgn(k^3)$.
Combining this property and the behaviors above, we see that the range where $f(l^0)>0$ is satisfied is
\begin{align} 
\begin{split}
\mf<l^0<l_-,~l_+<l^0& ~~(\text{for }s=- \sgn(k^3)),\\
\mf<l^0& ~~(\text{for }s=- \sgn(k^3)).
\end{split}
\end{align}

\section{Ward-Takahashi identity}
\label{app:WT}

We show that Bethe-Salpeter equation (\ref{eq:BS-2}) that is used to sum all the ladder diagrams is consistent with the Ward-Takahashi (WT) identity.
The identity for the vertex function reads~\cite{Aarts:2002tn}
\begin{align}
\label{eq:WT-id} 
p_\mu\varGamma^\mu(k+p,k)
&= [S^A(k_L)]^{-1}-[S^R(k_L+p_L)]^{-1},
\end{align}
where $\varGamma^\mu(k+p,k)$ is a vertex function where the two quarks have momenta $p+k$ and $k$, and $p$ is the momentum of the photon. 
This equation reduces to 
\begin{align} 
\label{eq:WT-id-smallp}
p_\mu\varGamma^\mu(k)
&= -2i{\text{Im}}\varSigma^R(k_L),
\end{align}
at $p\rightarrow 0$.

By multiplying Eq.~(\ref{eq:BS-2}) with $p_\mu$, we get
\begin{align} 
\begin{split}
p_\mu \varGamma^\mu(k)
&= \Slash{p} 
+g^2\Cf\int_l   \gamma_\alpha [S^{R}(l_L)-S^{A}(l_L)] \gamma_\beta  \rho^{\alpha\beta}_{D}(k-l) \\
&~~~ \times\left(\nf(-l^0)+\nb(k^0-l^0)\right) [R^f(\vk_\perp-\vl_\perp)]^2 ,
\end{split}
\end{align}
where we have used the WT identity (\ref{eq:WT-id}) in the right-hand side.
By taking the limit $p\rightarrow 0$, the right-hand side becomes
\begin{align} 
\begin{split}
&ig^2\Cf\int_l   \gamma_\alpha (\Slash{l}_L+\mf){\cal P}_+\rho_S(l_L)
\gamma_\beta  \rho^{\alpha\beta}_{D}(k-l) \\
&~~~ \times\left(\nf(-l^0)+\nb(k^0-l^0)\right) [R^f(\vk_\perp-\vl_\perp)]^2,
\end{split}
\end{align}
which is found to be equal to $-2i{\text{Im}}\varSigma^R(k_L)$ by using Eq.~(\ref{eq:ImSigma-oneloop}).
This equation is none other than the WT identity in the small $p$ limit, Eq.~(\ref{eq:WT-id-smallp}), so we see that the Bethe-Salpeter equation is consistent with the WT identity.

\section{Equivalence to linearized Boltzmann equation}
\label{app:linear-Boltzmann}

We show that our summation scheme of the quark damping rate and the ladder diagrams is equivalent to the linearized Boltzmann equation in this Appendix.
We start with the Boltzmann equation in an electromagnetic field 
for the distribution function of the LLL quarks in the one spatial dimension~\cite{another-paper}:
\begin{align}
\label{eq:Boltzmann}
[\partial_T+v^3\partial_Z
+e\qf E^3(T,Z)\partial_{k^3}]n^f(k^3,T,Z)
&=C[n],
\end{align}
where $n^f(k^3,T,Z)$ is the distribution function for the quark with flavor index $f$ whose momentum is $k^3$ and space-time position is $(T,Z)$, $v^3\equiv \partial \epsilon^L_k/(\partial k^3)=k^3/\epsilon^L_k$, and $C[n]$ is the collision integral, whose expression will be given later.

We linearize the distribution function in terms of $\vE$ as $n^f(k^3,T,Z)=\nf(\epsilon^L_k)+\delta n^f(k^3,T,Z)$.
Then, the linearized version of Eq.~(\ref{eq:Boltzmann}) reads
\begin{align} 
\label{eq:linearized-Boltzmann}
\begin{split}
e\qf  E^3(T,Z)
\beta v^3\nf(\epsilon^L_k)[1-\nf(\epsilon^L_k)]
&=
C[\delta n^f(k^3,T,Z) ],
\end{split}
\end{align}
Here, we consider the case that the electric field is constant and homogeneous, 
so that $\delta n^f$ does not depend on $T$ and $Z$.
The induced current is given by the distribution function as 
\begin{align}
\label{eq:current-Boltzmann}
\begin{split}
j^3(T,Z)&= 2e\sum_f \qf \Nc \frac{|\Bf|}{2\pi}
\int\frac{dk^3}{2\pi} v^3 \delta n^f(k^3,T,Z),
\end{split}
\end{align}
where we have taken into account the color, Landau, and quark/anti-quark degeneracies. 

First, we examine the relaxation time approximation, $C[\delta n^f]\simeq -\tau^{-1}_k \delta n^f$ 
with $\tau_k$ being the relaxation time. Then, the current reads 
\begin{align}
\begin{split}
j^3&= e^2\sum_f \qf^2 \Nc \frac{|\Bf|}{2\pi}
4 \beta E^3 \int^\infty_0 \frac{dk^3}{2\pi} (v^3)^2 \tau_k \\
&~~~\times \nf(\epsilon^L_k)[1-\nf(\epsilon^L_k)]\\
&=e^2\sum_f \qf^2 \Nc \frac{|\Bf|}{2\pi}
\frac{2}{\pi} \beta E^3 \int^\infty_{\mf} dk^0 v^3 \tau_k \\
&~~~\times \nf(k^0)[1-\nf(k^0)].
\end{split}
\end{align}
By using $j^3=\sigma^{33}E^3$, we get
\begin{align}
\begin{split}
\sigma^{33}
&=e^2\sum_f \qf^2 \Nc \frac{|\Bf|}{2\pi}
\frac{2}{\pi} \beta \int^\infty_{\mf} dk^0 \frac{\sqrt{(k^0)^2-\mf^2}}{k^0} \tau_k \\
&~~~\times \nf(k^0)[1-\nf(k^0)].
\end{split}
\end{align}
By comparing this expression with Eq.~(\ref{eq:sigma33-result-LL}), we see that the diagrammatic result at the one-loop order agrees with the result obtained from the Boltzmann equation in the relaxation time approximation, if we identify $\xi_k= \tau^{-1}_k/2$. 

Let us go beyond the relaxation time approximation, and evaluate the collision integral.
For the 1 to 2 scattering, the collision integral is given by 
\begin{align}
\label{eq:collision-integral}
\begin{split}
C[n]
&=\frac{1}{2\epsilon^L_k}\int_l
\rho_D(k+l)\rho_S(l_L) |M|^2 [R^f(\vk_\perp+\vl_\perp)]^2 \\
&~~~\times\{\nb(\epsilon^L_k+l^0)[1-n^f(k^3)][1-n^f(l^3)] \\
&~~~-[1+\nb(\epsilon^L_k+l)]n^f(k^3) n^f(l^3)
\},
\end{split}
\end{align}
where the matrix element is given by 
\begin{align}
\begin{split}
|M|^2&= g^2\Cf P^\parallel_{\mu\nu}(k+l)
\Tr[(\Slash{l}_L-\mf)\gamma^\mu {\cal P}_+ (\Slash{k}_L+\mf)\gamma^\nu] \\
&=4g^2\Cf \mf^2.
\end{split}
\end{align}
The linearized version of Eq.~(\ref{eq:collision-integral}) is
\begin{align}
\label{eq:linear-collision-integral}
\begin{split}
C[\delta n^f] 
&=-\frac{2g^2\Cf \mf^2}{\epsilon^L_k}
\int_l \rho_D(k+l)\rho_S(l_L)  [R^f(\vk_\perp+\vl_\perp)]^2 \\
&~~~\times\{\delta n^f(k^3)[\nb(\epsilon^L_k+l^0)+\nf(l^0)] \\
&~~~-\delta n^f(l^3)[\nb(\epsilon^L_k+l^0)+\nf(\epsilon^L_k)],
\end{split}
\end{align}
where we note that $\delta n^f(l^3)$ has a minus sign because the deviation of the anti-quark distribution function from equilibrium value has an opposite sign compared with that of quark.
The first term becomes $ -\tau^{-1}_k \delta n^f(k^3)$ in the relaxation time approximation.
By comparing this expression with Eq.~(\ref{eq:damping-tracedone}), we see that it reproduces the result of $\xi_k$ in the diagrammatic calculation.

By introducing $W_k$ that satisfies $\delta n^f(k^3)=e\qf \beta\nf(\epsilon^L_k)[1-\nf(\epsilon^L_k)]W_k E^3/2$, Eq.~(\ref{eq:linearized-Boltzmann}) whose left-hand side is replaced with Eq.~(\ref{eq:linear-collision-integral}) becomes
\begin{align} 
\label{eq:Boltzmann-inteq}
\begin{split} 
&g^2\Cf \mf^2
\int_l \rho_D(k+l)\rho_S(l_L)  [R^f(\vk_\perp+\vl_\perp)]^2 \\
&~~~\times
[W_k-W_l] 
[\nb(\epsilon^L_k+l^0)+\nf(l^0)] 
= k^3 , 
\end{split}
\end{align}
where we have used $\nf(l^0)[1-\nf(l^0)]
[\nb(\epsilon^L_k+l^0)+\nf(\epsilon^L_k)]= \nf(\epsilon^L_k)[1-\nf(\epsilon^L_k)] 
[\nb(\epsilon^L_k+l^0)+\nf(l^0)] $.
The relation to the current is, by using Eq.~(\ref{eq:current-Boltzmann}),
\begin{align}
\begin{split}
j^3&= e^2\sum_f \qf^2 \Nc \frac{|\Bf|}{2\pi} E^3 \beta
\int^\infty_{\mf}\frac{dk^0}{\pi} \nf(k^0)[1-\nf(k^0)]W_k .
\end{split}
\end{align}
By comparing this expression with Eqs.~(\ref{eq:Pi-12-G-aarr}) and (\ref{eq:Gaarr-A}), and Eq.~(\ref{eq:Boltzmann-inteq}) with Eq.~(\ref{eq:BS-4}), we see that solving the linearized Boltzmann equation taking into account the full collision integral corresponds to the summation of the ladder diagram in diagrammatic analysis, by identifying $W_k=A^3(k)$.


\end{document}